\documentclass[aps,twocolumn,showpacs,superscriptaddress,prc]{revtex4}

\usepackage{amsmath}
\usepackage{graphicx}
\DeclareGraphicsExtensions{.eps,.jpg}

\begin{document}
\newcommand{\tb}{ {\bf {t}}}

\newcommand {\Co}{$^{57}$Co}

\newcommand {\Ar}{$^{39}$Ar}

\newcommand {\Na}{$^{22}$Na}

\newcommand {\Kr}{$^{83}$Kr$^{\mathrm{m}}$}

\newcommand {\mus}{$\mu$s}

\newcommand{\leff}{$\mathcal{L}_\textrm{eff}$}

\newcommand {\fP}{$f_{p}$}

\title{Scintillation yield and time dependence from electronic and nuclear
  recoils in liquid neon}
\author{W.~H.~Lippincott}
\email{hugh@fnal.gov}
\affiliation{Fermi National Accelerator Laboratory, Batavia, Illinois 60510}
\affiliation{Department of Physics, Yale University, New Haven, Connecticut 06511}
\author{K.~J.~Coakley}
\affiliation{National Institute of Standards and Technology, Boulder, Colorado
  80305}
\author{D.~Gastler}
\affiliation{Department of Physics, Boston University, Boston, Massachusetts 02215}
\author{E.~Kearns}
\affiliation{Department of Physics, Boston University, Boston, Massachusetts 02215}
\author{D.~N.~McKinsey}
\affiliation{Department of Physics, Yale University, New Haven, Connecticut 06511}
\author{J.~A.~Nikkel}
\affiliation{Department of Physics, Yale University, New Haven, Connecticut 06511}

\date{\today}

\begin{abstract}
We have performed measurements of scintillation light in liquid neon, observing a signal yield in our detector as high as (3.5 $\pm$ 0.4) photoelectrons/keV. We measure pulse shape discrimination efficiency between electronic and nuclear recoils in liquid neon from 50 and 300 keV nuclear recoil energy. We also measure the \leff\, parameter in liquid neon between 30 and 370 keV nuclear recoil energy, observing an average  \leff$=0.24$ above 50 keV. We observe a dependence of the scintillation time distribution and signal yield on the pressure and temperature of the liquid neon. 
\end{abstract}


\maketitle

  \section{Introduction}
  \label{sec:intro}
The DEAP/CLEAN collaboration has proposed to use liquid argon and liquid neon as scintillating targets in large scale detectors to observe both $pp$-solar neutrinos and
dark matter in the form of weakly interacting massive particles (WIMPs)~\cite{McKinsey:2000,Boulay:2005,McKinsey:2005b}. In particular, the MiniCLEAN experiment is designed to do a dark matter search with first liquid argon and then liquid neon as both a systematic check on backgrounds and as a precursor to a multi-tonne scale  detector~\cite{McKinsey:2007}. Because of the expected $A^2$-dependence of the cross section for WIMP-nucleon scattering~\cite{Jungman:1996}, filling the same detector alternately with argon and neon will change the expected dark matter signal in a known way while maintaining an identical level of external backgrounds. Here, we extend previous studies of the scintillation properties of liquid neon~\cite{Nikkel:2007}.

Like the
other liquefied noble gases, liquid neon is relatively inexpensive, is easily purified
of radioactive contaminants, scintillates brightly when exposed to ionizing radiation and is dense enough to self-shield,
reducing the background level in the center of a larger volume of
liquid.  The key for
dark matter detection is to be able to suppress the electronic recoils that
make up most of the backgrounds from the nuclear recoils that would make up a
WIMP signal by use of some combination of self-shielding and discrimination. Because $\nu_e-e$ scattering also produces electronic
recoils, discrimination does not improve the performance of a neutrino detector. However, neon has no
long-lived radioactive isotopes and is more
readily purified of such contaminants as $^{39}$Ar and
$^{85}$Kr~\cite{Harrison:2006} than the heavier noble gases, rendering it an
ideal target for low-energy neutrinos. 

When ionizing radiation interacts in liquid neon or any other liquefied noble gas, ultraviolet light
is produced via scintillation. The incoming radiation collides with an electron
or nucleus in the liquid and deposits energy. The resulting electronic or
nuclear recoil then excites or ionizes other neon atoms. The excited atoms quickly combine with surrounding ground state atoms to form dimer
states, Ne$_2^*$. The ions also bond with ground state atoms to form
ionized molecules, Ne$_2^+$, which can in turn recombine with free electrons to also form the excited dimer states.
Finally, the dimer states decay emitting scintillation photons with wavelength 80 nm. Because these photons have insufficient energy to excite neon atoms~\cite{Packard:1970}, the liquid
does not absorb its own scintillation light and large light yields can be attained from liquid noble gas detectors.

The metastable
molecules are produced in both singlet and triplet states. The singlet
state decays in nanoseconds, while the triplet state undergoes a forbidden spin
flip before decaying, extending the triplet lifetime to 15~\mus\, in neon~\cite{Nikkel:2007}. Because nuclear and electronic recoils produce
different ratios of singlets to triplets, pulse timing can be used to
identify the source of the initial excitation via pulse shape discrimination
(PSD)~\cite{McKinsey:2005b,Boulay:2005}.  As the PSD depends directly on the scintillation timing, we have attempted to re-measure the time constants of liquid neon scintillation in this work. We discovered two additional time components in the light signals produced by ionizing radiation, and we also found that both the intensity and lifetime of each component changes significantly with the temperature and pressure of the liquid, a new feature that is not observed in the heavier noble gases. 

A second property of scintillation in liquefied noble gases is that nuclear recoils produce less light
than electronic recoils of the same energy. The ratio of signal yields at zero electric field for the
two event classes
is known as the nuclear recoil scintillation efficiency, or \leff. The value
of \leff\, in combination with the discrimination efficiency sets the nuclear recoil
energy analysis threshold of a liquefied noble gas dark matter detector. Because in general  
$\mathcal{L}_\textrm{eff} < 1$, two energy
scales may be employed, denoted ``keVee'' and ``keVr.'' The actual energy deposited by a nuclear recoil is expressed in units of keV recoil or
keVr. Most calibration sources used in liquid noble gas detectors are
$\gamma$ or $\beta$ sources that produce electronic recoils in the
liquid. Therefore, the energy calibration gleaned from these sources refers to the
energy deposited by an electronic recoil, or keV electron
equivalent (keVee). In practice, we often have no direct
energy calibration for nuclear recoils and so use electron
equivalent energies. The conversion factor is \leff. At zero electric field
\begin{equation}
E [\mathrm{keVee}] = E [\mathrm{keVr}] \times \mathcal{L}_\mathrm{eff}.
\end{equation}

PSD has been studied in
argon~\cite{Lippincott:2008,Boulay:2009,Benetti:2007} and
xenon~\cite{Dawson:2005,Kwong:2010,Ueshima:2011}. Our group performed a measurement of PSD in neon as
well, but nuclear recoils were observed at only one
energy~\cite{Nikkel:2007}. Similarly, \leff\, has been measured in argon~\cite{Gastler:2009} and
xenon~\cite{Chepel:2006,Aprile:2005,Aprile:2009,Manzur:2010,Plante:2011}, but at only one
energy in neon. In this paper, we report measurements of scintillation light
in liquid neon over a wide energy range. We measure PSD from 50 to 300 keVr and \leff\, from 30 to 370 keVr . We also present results on the time
dependence of liquid neon scintillation light as a function of temperature.

  \section{Experimental Apparatus}
  \label{sec:det_design}

\begin{figure}[htbp]
 \begin{center}
  \includegraphics[width=7.5cm]{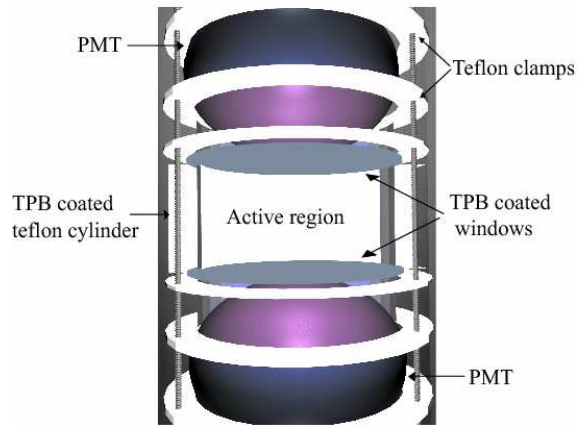}
  \end{center}
          \caption[Scintillation cell]
                  {(Color online) Schematic representation of the MicroCLEAN scintillation cell.}
          \label{fig:cell}
\end{figure}

The apparatus used in these measurements is named
MicroCLEAN and has been described in detail
elsewhere~\cite{Lippincott:2008}. A schematic is shown in Fig.~\ref{fig:cell}.
The detector has an active volume of 3.14 liters viewed by
two 200-mm-diameter Hamamatsu R5912-02MOD photomultiplier tubes
(PMTs), which are specifically designed for use in cryogenic
liquids. The active volume
is defined by a polytetrafluoroethylene (PTFE) cylinder 200~mm in diameter and 100~mm in height, with
two 3-mm-thick fused-silica windows at top and bottom.  All inner surfaces of the
PTFE and windows are coated with $(0.20 \pm 0.01)$ mg/cm$^2$ of tetraphenyl butadiene (TPB)~\cite{McKinsey:1997},
which shifts the wavelength of the ultraviolet light to approximately 440~nm. 
The
TPB thickness on the PTFE cylinder was reduced by $33\%$ relative
to that used in~\cite{Lippincott:2008} to mitigate TPB absorption of the wavelength-shifted
light. The MicroCLEAN vessel is contained within a stainless steel vacuum
dewar for thermal insulation purposes.

\begin{figure}[htbp]
\begin{center}
  \includegraphics[width=7.5cm]{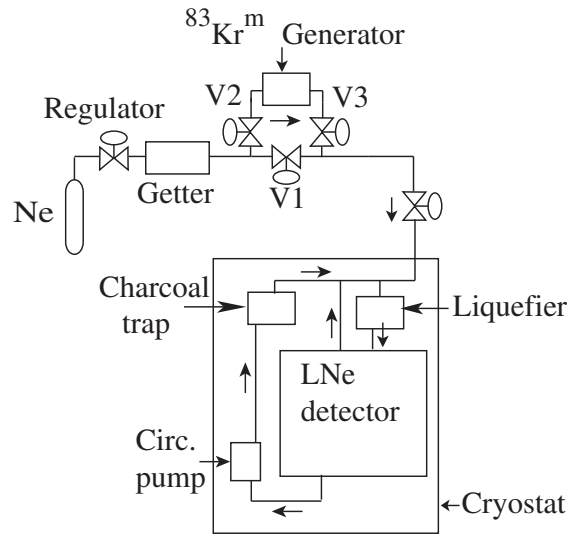}
  \end{center}
          \caption[Gas handling system]
                  {The gas handling system for the MicroCLEAN runs described
                    in this paper. The circulation pump consists of a heater
                    that evaporates the liquid; the gas then flows through the
                  charcoal trap before being reliquefied. }
          \label{fig:gashandle}
          
\end{figure}

The gas handling system is shown schematically in Fig.~\ref{fig:gashandle}.   Ultra-high-purity neon flows through
a Nupure Omni 1000~\cite{Nupure}\footnote{Commercial equipment, instruments, or materials
  are identified in this paper to adequately specify the experimental
  procedure. Such identification implies no recommendation or endorsement by
  NIST, nor does it imply that the materials or equipment identified are
  necessarily the best available for the purpose.} getter and an external charcoal trap at 77 K (not shown) before passing into a vacuum cryostat
through a tube at the top. Previous studies have shown that activated charcoal provides a very effective purifier of many impurities in neon, including radon~\cite{Harrison:2006}. Once inside the vacuum cryostat, the gas flows into a copper liquefier cell
attached to the second stage of a Cryomech Model PT805 cryorefrigerator before
dripping into the MicroCLEAN vessel. 

There is a
circulation loop inside the vacuum dewar for continuous purification of the
neon, again using activated charcoal as the purifier. Liquid neon
flows out of the bottom of the MicroCLEAN vessel into a
nearby plumbing volume containing a heater. The heater (labeled ``Circ. pump'' in Fig.~\ref{fig:gashandle}) acts as a pump by boiling
the liquid; the resulting pressure differential between the heater and the liquefier drives gas up a tube through a charcoal trap mounted inside the cryostat before reentering the top of the
liquefier, where gravity pulls the liquid back into the main chamber. 

The ``\Kr\, Generator'' shown in Fig.~\ref{fig:gashandle}
refers to the setup for introducing \Kr\, into the liquid neon for calibration
purposes as described in Sec.~\ref{sec:Calibrations} and~\cite{Lippincott:2010}.  The \Kr\, generator is connected to the gas inlet line just outside the vacuum
dewar.  The incoming gas can be diverted through the trap on its way into the
detector to introduce \Kr\, atoms into the neon. 

\subsection{Data acquisition and processing}
\label{sec:DAQ}
\begin{figure}[htbp]
\begin{center}
  \includegraphics[width=7.5cm]{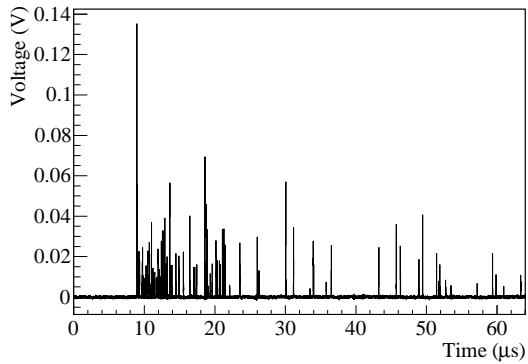}
\end{center}
         \caption
                  {Sample event from a single PMT digitized by the CAEN
                    V1720 waveform digitizer. More than eight microseconds of presamples are recorded to
                    measure the baseline. }
          \label{fig:sampletrace}
        \end{figure}

The data acquisition (DAQ) system consists of a 250 MHz, 12-bit CAEN V1720 waveform
digitizer (WFD). The two PMT channels are recorded with a record length
of 64~\mus~to collect as much of the long-lived triplet light as possible, and
a sample trace is shown in Fig.~\ref{fig:sampletrace}. The PMT signals are
passed through a LeCroy Model 612AM dual-output amplifier, with one copy
passed to the WFD and the other copy sent to triggering electronics. The
trigger is set in several different configurations. In most cases,
a trigger is generated when each PMT signal crosses a threshold of 1/2 times
the mean height of a
single photoelectron within a 100 ns coincidence window. For some low energy runs
when an external detector is used as a coincidence tag, the trigger is
generated when the sum of the two PMT signals crosses a threshold of roughly
1/2 the size of a photoelectron in coincidence with a signal in the
external tagging detector. In each 64~\mus~data record, approximately
8~\mus~of presamples are collected to measure the baseline of each pulse. The
data collected by the DAQ software are saved in a ROOT-based file
structure for analysis~\cite{Brun:1997}.

The data processing is similar to that described
in~\cite{Lippincott:2008}. The presamples are averaged to obtain a baseline
and a baseline root-mean-square, and the baseline is subtracted from each
trace. We integrate each trace from 50 ns before the trigger (defined as the time at which the voltage first rises above $20\%$ of its maximum
value) to 44~\mus~after the trigger to determine the total area of each pulse. In
performing this integral, we restrict the range of
integration to 100 ns windows around locations where the trace voltage
crosses a threshold of three times the baseline root-mean-square. This
integration method is a hybrid of single photoelectron counting and pulse
integration. We reject data for which the trigger times of the two PMTs
 differ by greater than 20 ns. We define an asymmetry parameter, 
\begin{equation}
A = \frac{S_T - S_B}{S_T + S_B},
\end{equation}
where $S_T$ and $S_B$ are the signal areas in the top and bottom PMTs. We
reject data for which the asymmetry parameter is farther than $2\sigma$ from
the mean value as determined by fitting a Gaussian function to the asymmetry data. 

As the triplet component of liquid neon scintillation is spread out over tens
of microseconds, there are many single photoelectron pulses well separated in
time in the tails of events. To obtain a single photoelectron spectrum for
each PMT, the last 30 \mus~of each trace is evenly divided into 150 ns
windows. If the voltage spread in the first and last 30 ns of each 150 ns
window is less than three times the size of the baseline root-mean-square,
the peak height and pulse area of the window are collected in a 2-D histogram; this restriction
prevents inclusion of partial single photoelectrons in the data set. A
Gaussian is fit to the 1-D histogram of the pulse areas to determine the mean
single photoelectron size. 

In previous tests of the R5912-02MOD, it was found
that the gain dropped substantially at 30 K~\cite{Nikkel:2007a}. For
the data presented here, we observe a sharp drop in PMT gain upon cooling to
liquid neon temperatures, followed by a slow decrease in gain over several
weeks before finally reaching a plateau. The low resulting gain  necessitates the application of
higher voltages to the PMTs and amplification from the LeCroy 610AM
amplifier;  for most of the data presented here,
the PMTs are biased at 1750 V, the absolute gain of the PMTs is $\approx1.1 \times
10^6$, and the PMT signals are further amplified by a factor of 10 with the LeCroy
amplifier.  A single photoelectron peak is resolved in the bottom PMT but not in the top PMT. To improve PMT performance, we use two high voltage lines to bias
the PMTs, with a separate high voltage running directly to the first dynode. By
tuning the first dynode voltage relative to the total voltage, we can achieve some
improvement in the single photoelectron spectra but not enough to recover a resolvable
single photoelectron peak in the top PMT.

\subsection{Radioactive sources}
\label{sec:Calibrations}
We use several radioactive sources for calibration purposes: the
\Kr\, generator described in~\cite{Lippincott:2010}, a 10-$\mu$Ci
\Na\, source for 511-keV and 1275-keV gamma rays, and a D-D neutron generator producing a
monoenergetic beam of 2.8-MeV neutrons~\cite{Thermo}. The \Kr\, generator
produces metastable \Kr\, nuclei that decay via two electromagnetic
transitions with energy 32.1 keV and 9.4 keV. The second transitions has a half life of
154~ns, and given the long time scale of liquid neon scintillation, both transitions are observed together as a
single waveform of total energy 41.5 keV.  The \Na\, source produces positrons that
immediately annihilate into two 511-keV gamma
rays with opposite momenta; we tag the second gamma ray using a NaI crystal
placed back-to-back with our apparatus to improve the data quality from the
\Na\, source and reduce backgrounds. We reject data where the trigger times in the NaI crystal and
the liquid neon differ by more than 30 ns, and we also make a cut to select events within 2$\sigma$ of the 
511-keV peak of the NaI crystal. 

\begin{figure}[htbp]
\begin{center}
 \includegraphics[width=80mm]{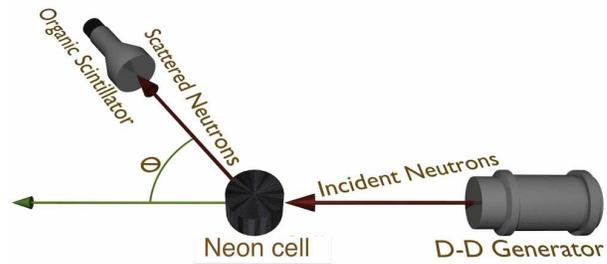}
\end{center}
          \caption{(Color online) Schematic of the neutron scattering setup.  
                  }
          \label{fig:nuc_scatt}
\end{figure}

We also use a tagging detector when investigating nuclear recoils with the
neutron generator. We require a PMT viewing BC501A organic scintillator to
record an event within 200 ns of an event in the liquid neon. The organic scintillator is contained in a cylinder with a diameter of 127 mm and a depth of 127 mm. We also make a PSD
cut in the organic scintillator and a cut on the time-of-flight (TOF) between the
event in the liquid neon and the tagging detector. These cuts will be discussed in more detail in Sec.~\ref{sec:LEff}.  Both the
neutron generator and the organic scintillator are located approximately 1.63~m from
MicroCLEAN, and the experimental setup can be seen schematically in
Fig.~\ref{fig:nuc_scatt}.  By changing the scattering angle $\theta$, we can choose the energy of the nuclear recoils that scatter just once in the liquid, $E_{rec}$, using simple kinematics:
\begin{eqnarray}\label{eq:NeutronScattering}
 \nonumber E_{rec} =&
\frac{2E_{in}}{(1+M)^2}[1+M-\cos^2(\theta) \\
&-\cos(\theta)\sqrt{M^2+\cos^2(\theta)-1}],
\end{eqnarray}
where $E_{in}$ is the incident neutron energy and $M$ is the atomic mass of the
target. We measure the angle by running a string marked at both ends and at its midpoint   from the generator around the dewar and back to the generator. We mark the location of the midpoint of the string on the dewar and then repeat the process for the organic scintillator. The ratio of the arc length between the two points on the dewar to the circumference (175.6 cm) provides the angle, $\theta$. The accuracy is limited by our knowledge of the midpoint of the string, and by repeated measurements we estimate the uncertainty on the arc length to be 6.4 mm, for an uncertainty on the angle of 1.3 degrees.

\subsection{Energy calibration}

We obtain an energy calibration by comparing $^{22}$Na data with a simulation of the detector performed with the Reactor
Analysis Tool (RAT)~\cite{RAT}. Combining Geant4 and ROOT~\cite{Agostinelli:2003,Brun:1997}
in a single simulation package, the RAT simulation of MicroCLEAN contains a complete optical
model
of the inner detector and PMTs and has been used previously in studies of
liquid argon scintillation~\cite{Lippincott:2008,Gastler:2009}. In general,
there are two features in the $^{22}$Na data that can be used for calibration
purposes: a 511-keV photoabsorption ``hump'' and the Compton edge. 

\begin{figure}[htbp]
\begin{center}
  \includegraphics[width=80mm]{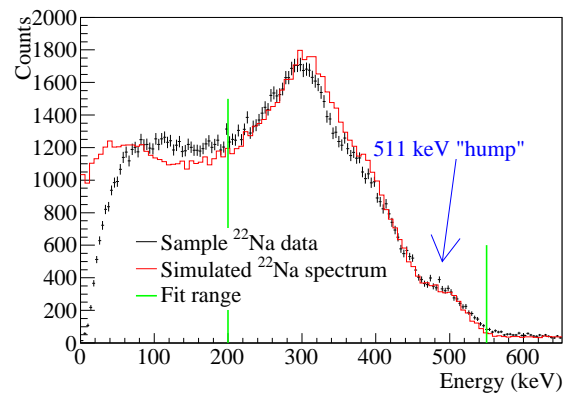}
\end{center}
\caption[Obtaining the energy calibration by data to simulation]{(Color
  online) The energy calibration is obtained by fitting the Compton edge and
  photoabsorption hump produced by 511 keV gamma rays to a simulation of the
  detector. The green lines denote the fit range. We
   currently have no satisfactory explanation for  the
excess of events between 40 keV and 200 keV in comparing the data to the simulation.}
\label{fig:LYSim}
\end{figure}

The
analysis begins by first obtaining a rough estimate for the signal yield from
the 511 keV hump, without using the simulation. We tune the parameters of
the simulation to achieve the estimated light yield before simulating the
$^{22}$Na source including a flat background contribution to account for the events observed above 600 keV from accidental coincidence of background events in the argon and 511 keV gamma rays in the NaI detector. We then re-fit the
data to the simulated spectrum, and the scale factor between the integrated charge from both PMTs to the energy scale of the simulation provides the energy
calibration; an example of this fit is shown in Fig.~\ref{fig:LYSim}. While the energy resolution is not an explicit
parameter of the fit, we do smear the photon counts from the simulated
spectrum with a Gaussian kernel having a standard deviation of $\sqrt{2\times\langle N_{pe} \rangle}$,
as this smearing provides better agreement between simulation and data. This
additional smearing suggests a source of noise in the real experiment that is
not being accurately modeled in the simulation, and a likely source of this
noise is the poor gain dispersion characteristics of the PMTs. It could also
be a property of the liquid neon.

The simulation does not accurately model the data below about 200 keV, predicting about $10\%$ fewer counts than observed down to the low energies where thresholding effects begin (about 40 keV for the run shown in Fig.~\ref{fig:LYSim}). This data excess is seen for a variety of different gain and trigger threshold settings, and it is currently unexplained, although asymmetry effects between the PMTs smearing
out very low energy events is one possibility, as is an additional source of background in the detector.  Given the discrepancy between data and simulation and the fact that that the trigger and gain settings were varied throughout the course of the run, the energy calibration fit is performed between
200 and 550 keV.  

There is some tension between the best fit at the Compton edge and the 511 keV hump, and the average $\chi^2/$NDF is 2.5 for all runs with the nominal fit range. If the fit is restricted to the range between 350 keV and 550 keV, the agreement is much better, with an average $\chi^2/$NDF of 1.4. We estimate the systematic uncertainty caused
by this tension to be $3\%$ by fitting each feature independently and taking the difference in the resulting energy scale relative to the nominal fit as the uncertainty.  Performing the fit without the flat background contribution in the simulation changes the calibration by less than $1\%$.
An additional source of uncertainty in this calibration are the statistics
of fitting the simulated and observed spectra, but this 
source of uncertainty is estimated to be $1\%$.  
 
\begin{figure}[htbp]
\begin{center}
  \includegraphics[width=80mm]{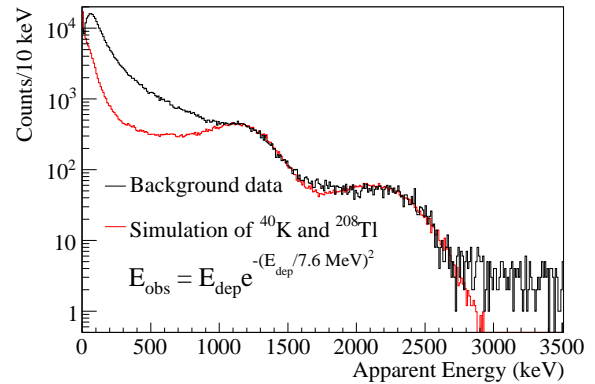}
\end{center}
\caption{(Color
  online) We check the energy calibration obtained from the $^{22}$Na source by comparing background data to simulations containing $^{40}$K and $^{208}$Tl components. We clearly see the two Compton edges of the 1.4 MeV and 2.6 MeV gammas produced by these sources. We do not attempt to simulate any other backgrounds, so the divergence of data and MC below 1000 keV is not surprising. There is evidence for a non-linearity in the energy scale at these high energies that is likely related to a  the response of the top PMT, and we correct for this using Eq.~\ref{eq:Energy}.}
\label{fig:BGComp}
\end{figure}

We cross check the energy calibration obtained from the $^{22}$Na source with both higher and lower energy references.  For our lower energy reference, we use the \Kr\, source described in the next section which is the primary reason we believe the energy calibration is accurate to within our uncertainties. For the high energy reference, we look at the Compton edges of the 1.4 MeV and 2.6 MeV gamma rays produced in the decay of $^{40}$K and $^{208}$Tl, as there are a number of these events in background data. Figure~\ref{fig:BGComp} shows background data with the energy scale derived from the $^{22}$Na calibration compared to a simulation of 1.4 MeV and 2.6 MeV gamma rays using the RAT simulation. As the purpose of this simulation is solely to cross-check the accuracy of the energy scale derived from the $^{22}$Na at higher energies, we do not attempt to simulate any other sources of background (i.e. lower energy gamma rays in the detector materials, cosmic rays, or other backgrounds); therefore, we do not expect the simulation to agree with the data below 1 MeV. We find evidence for a slight non-linearity in either the energy response of the liquid neon or the PMTs at these energies, as the background data do not agree with the simulation if we assume a linear response of the detector up to 2.6 MeV.  The fit is improved if the observed energy (as calibrated by the $^{22}$Na source) is actually representing the following function of the deposited energy in the simulation:
\begin{equation}
\label{eq:Energy}
\langle E_{obs}\rangle = E_{dep}\times e^{-(E_{dep}/7.6\,\mathrm{MeV})^2}.
\end{equation}
It is likely that this dependence is caused by a non-linearity of the top PMT, as a comparison using just the signal in the bottom PMT requires no energy scaling to provide agreement. We also find that the data are best fit with an energy resolution of $\sigma = 2.4\sqrt{E_{dep}}$. Between 1 MeV and 2.7 MeV, we find a $\chi^2/$NDF for this fit of 1.27. 

\subsection{The \Kr\, source}

\begin{figure}[htbp]
\begin{center}
  \includegraphics[width=80mm]{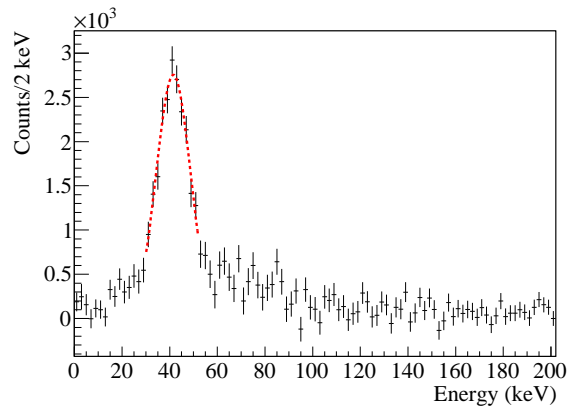}
\end{center}
\caption[Energy spectrum of \Kr\, in liquid neon]{(Color online) Energy spectrum of a background subtracted \Kr\ run in
                    neon. As discussed in the text, the observed light
                    yield for these data is $(3.0 \pm 0.3)$
                    photoelectrons/keVee. The resolution is $19\%$ ($\sigma$/E of the Gaussian fit shown in red) at 41.5 keV.}
\label{fig:Kr}
\end{figure}

Previously, we reported on use of a \Kr\, generator as a source of low energy electronic recoils in liquid argon and neon~\cite{Lippincott:2010}. This Kr\, generator became available at the end of the experiment described here to provide a
second, independent energy calibration at 41.5 keV, with a clear peak
appearing in the data as shown in
Fig.~\ref{fig:Kr}. When the energy calibration is derived from the $^{22}$Na spectrum as described above, the energy of the \Kr\, peak agrees with 41.5 keV within statistical uncertainties, suggesting both that the energy calibration is accurate and that the response of the MicroCLEAN detector is linear between 9.4 keV and 511 keV (recall that the \Kr\, atoms emit two electrons at 9.4 keV and 32.1 keV to provide the full energy peak). The resolution of our detector at 41.5 keV is $19\%$ ($\sigma$/E of the Gaussian fit shown in Fig.~\ref{fig:Kr}).

\subsection{Signal yield}

The energy calibration described above uses the total signal from both PMTs. Given the absence of a well-defined mean single photoelectron in the top PMT,
the absolute signal yield is estimated from the bottom PMT alone by matching the energy calibration to the integrated charge of the bottom PMT divided by the size of the single photoelectron of the bottom PMT. After ending the liquid neon run described in this report, we filled the detector with liquid argon without warming up to room temperature or opening the system; at the higher temperatures of liquid argon, we were able to resolve the single photoelectron response of both PMTs. From these measurements, we determined that
the top tube observes $(6 \pm 3)\%$ more signal than the bottom
tube~\cite{Lippincott:2010}, due to some combination of photon collection efficiency and photocathode quantum efficiency. Therefore, we extrapolate the total light yield
in the liquid neon to include both tubes based on the observed PMT efficiencies.  Using this method, 
we determine a photoelectron yield of $(3.0 \pm 0.3)$
photoelectrons/keVee from the \Kr\, data.
The total uncertainty includes an estimate of the uncertainty in the relative efficiency of the two tubes introduced by changing the temperature from 85 K to 25 K, but it is dominated by the single photoelectron
response of the bottom PMT. 

We next use the \Kr\, data to validate the $^{22}$Na calibration method. Neglecting the systematic uncertainties from the PMT photoelectron response and efficiency, as they apply equally to both methods
for determining the light yield, the standard $^{22}$Na energy calibration taken at the same time as the \Kr\, data returns 
 a light yield of $(3.00 \pm 0.02)$ photoelectrons/keVee, in agreement with 
 the \Kr\, calibration. This agreement provides confidence that we can trust the $^{22}$Na calibration for the data taken before the the \Kr\, source became available.
 
   As will be discussed in Sec.~\ref{sec:DetStab}, the
signal yield was not stable throughout the course of the run; the largest
signal yield observed in this study was $(3.5\pm0.4)$ photoelectrons/keVee, a factor of about 3.5 larger than we achieved in our previous measurement (0.9 photoelectrons/keVee~\cite{Nikkel:2007}). We attribute the increase to immersing the PMTs in the liquid instead of viewing through pressure windows, improved photocathode coverage, and a change in the thickness of the wavelength shifter.

\subsection{Detector stability}\label{sec:DetStab}
\begin{figure}[htbp]
\begin{center}
  \includegraphics[width=80mm]{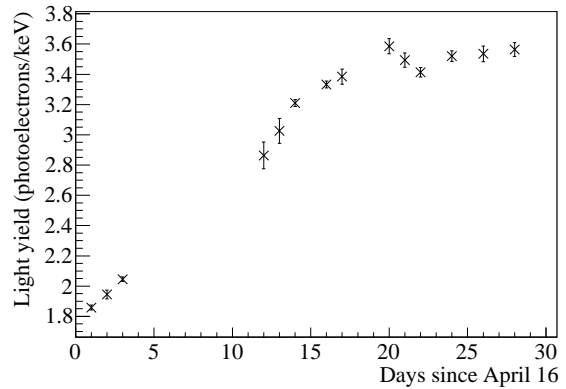}
\end{center}
\caption[Light yield versus time in MicroCLEAN during purification]{The light yield in the detector as a function of time
                    with the circulation pump to the charcoal trap
                    engaged for a liquid temperature of 28.7 K as determined by the $^{22}$Na source. The light yield reached a maximum at $(3.5\pm0.4)$
                    photoelectrons/keVee. The error bars are statistical only and do not include the uncertainty introduced by the single photoelectron response of the bottom PMT.}
\label{fig:Circulation}
\end{figure}

The presence of impurities can reduce the amount of light detected both by
quenching excimer states non-radiatively or by absorbing the VUV scintillation
light, and both processes have been studied in
liquid argon~\cite{Acciarri:2008,Acciarri:2008a,Himi:1982}. We monitored both the light yield and the triplet lifetime over the course of the
experiment to gauge the effect of impurities. Initially, the signal yield of
the detector was only 1.9 photoelectrons/keVee. We then engaged the circulation
pump and flowed neon through the charcoal trap shown in the schematic of Fig.~\ref{fig:gashandle} to remove impurities. The light
yield subsequently increased, reaching a peak at $(3.5\pm0.4)$
photoelectrons/keVee  as shown in
Fig.~\ref{fig:Circulation}, providing evidence that impurities were being contained
by the charcoal.  Because we could not purge the internal charcoal trap
while running the experiment, it is not clear whether all the impurities in
the neon were removed or whether the trap became saturated. These data were taken for a liquid temperature of 28.7 K. 

Both the light yield and the triplet lifetime
varied with time. As will be discussed in
Sec.~\ref{sec:Temp}, both of these
parameters are strongly and unexpectedly affected by the temperature and pressure of the
liquid neon. To stabilize the temperature and pressure, we disengaged the
circulation pump for the remainder of the run, potentially allowing
contaminants to build up in the neon and decreasing the observed light yield in the
\Kr\, data taken at the very end of the experiment to the $(3.0 \pm 0.3)$ photoelectrons/keVee
reported above. 

To account for the temperature effects, the PSD and \leff\, data were collected at three different temperatures, 26.7 K, 27.8
K and 28.7 K, where each data set was taken over a roughly two week period. We measured the light yield approximately every other day during each run to determine the stability of the detector for each temperature set point. The observed signal yields were $2.74\pm0.03$,
$3.15\pm0.08$ and $3.13\pm0.03$ photoelectrons/keVee at 26.7 K, 27.8 K and 28.7 K respectively, where the error bars represent the
root-mean-square error of all the calibration runs taken at the operating temperature. We therefore conclude that the detector signal yield was stable to better than 3$\%$ for each temperature setting.

\section{Pulse Shape Discrimination}
\begin{figure}[htbp]
\begin{center}
  \includegraphics[width=80mm]{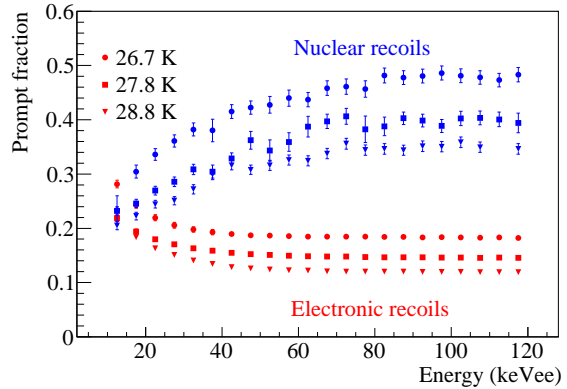}
\end{center}
\caption[Prompt fraction versus energy in liquid neon]{(Color online) Mean prompt fraction as a function of energy for electronic and
    nuclear recoils at three different temperatures. The error bars include
    both statistical and systematic uncertainties and are generally dominated by a
   $2.5\%$ uncertainty stemming from differences in the two PMTs.}
\label{fig:FP}
\end{figure}

We define the prompt fraction \fP\, as 
\begin{equation}
  f_{p} = \frac{\int_{T_i}^{\xi} V(t) dt}{\int_{T_i}^{T_f} V(t) dt},
  \label{eq:fp}
\end{equation}
where $V(t)$ is the voltage trace from the PMT, $\xi =220$ ns is an
integration period determined to optimize the PSD~\cite{Nikkel:2007}, $T_i = t_0 - 50$ ns, $T_f =
t_0 + 44$ \mus, and $t_0$ is the trigger time. As mentioned in the previous
section, we collected data at  26.7 K, 27.8 K and 28.8 K. In energy bins of 5 keVee, we fit a Gaussian function to the empirical
distributions to determine the mean prompt fraction, $\hat{f_p}$, for nuclear and
electronic recoils, and Figure~\ref{fig:FP} shows $\hat{f_p}$ as a
function of energy for both classes of events.  The error bars include both
statistical uncertainties and a systematic uncertainty of $2.5\%$ stemming from differences
in the two PMTs, with the systematic error generally dominating; the total uncertainty
is smaller than the size of the markers for the electronic recoil data. There is also a small systematic error associated with fitting an asymmetric distribution with a symmetric Gaussian function that we do not include in this analysis. The nuclear
recoil data were acquired in coincidence with a BC501A organic scintillator module as described in Sec.~\ref{sec:Calibrations}. Two effects may be observed: first, the mean prompt fractions for nuclear and electronic recoils converge at low energies. This effect has been observed in liquid argon~\cite{Lippincott:2008,Boulay:2009} and is likely due to the fact that $dE/dx$ for nuclear and electronic recoils also converge as the energy decreases. Also, the mean
prompt fraction decreases with increasing temperature and pressure for both
electronic and nuclear recoils at all energies. This will be discussed
further in Sec.~\ref{sec:Temp}.

\begin{figure}[htbp]
\begin{center}
  \includegraphics[width=80mm]{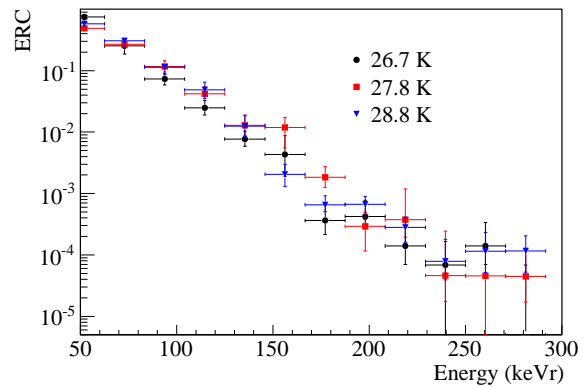}
\end{center}
\caption[PSD in liquid neon]{(Color online) ERC versus energy for the three data sets. The x-axis is given in relevant units for dark matter searches, keVr, by assuming \leff\ = 0.24 (see Sec.~\ref{sec:LEff}).}
\label{fig:PSD}
\end{figure}

We define the electronic recoil contamination (ERC) as the probability of
mistaking an electronic recoil for a nuclear recoil given a particular level of
nuclear recoil acceptance. We estimate the ERC as the
number of tagged electronic recoil events with $f_{p}>
\hat{f}_{p,\mathrm{nuclear}}$, setting the nuclear recoil acceptance level to approximately
$50\%$.  The observed ERC at each temperature is shown in
Fig.~\ref{fig:PSD}. Our observed ERC is about a factor of 5-10 better than
that observed in~\cite{Nikkel:2007}. Theoretically, we might have expected a
stronger improvement given that we collect at least three times as much light. The
smaller than expected improvement is in part attributable to the use of a
$\hat{f}_{p,\mathrm{nuclear}}$ that changes with energy, instead of the flat
cut value assumed previously in~\cite{Nikkel:2007}. As shown in Fig.~\ref{fig:FP}, the mean prompt
fractions for electronic and nuclear recoils converge at lower energies,
reducing the discrimination power. 

\begin{figure}[htbp]
\begin{center}
  \includegraphics[width=80mm]{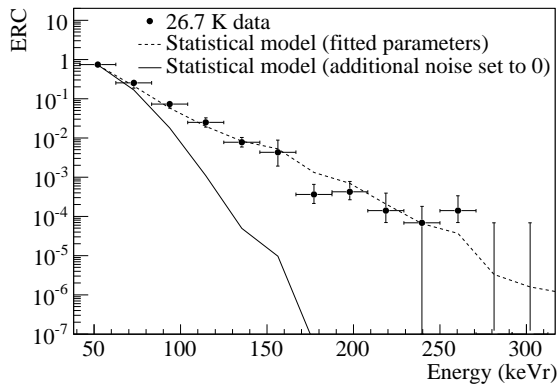}
\end{center}
\caption[PSD versus temperature with model prediction]{PSD for
  the 26.7 K data set individually, along with a fit to the statistical
    model. The solid line shows the model prediction with additional
    noise terms set to zero. We note that the effect of the additional noise is most relevant for energies above 100 keVr.}
\label{fig:PSDModel}
\end{figure}

A second cause of weak PSD is the
electronic noise in the experiment.  We fit the data using a statistical
model for \fP\, described
in~\cite{Boulay:2009,Lippincott:2008}, and Figure~\ref{fig:PSDModel} shows the
PSD for the 26.7 K data
 with a fit to this model. In~\cite{Lippincott:2008}, we described a method for estimating
the size of the additional noise for comparison between detectors, modeling
the prompt and late noise as a constant multiplied by the mean prompt and late
signals: $\sigma_p^2 = C_p\mu_p$ and $\sigma_l^2 = C_l\mu_l$. As discussed in ~\cite{Lippincott:2008}, we are wary of interpreting these fit parameters as a literal measurement of the noise; however, we do believe they can be used as points of comparison. With that
method, the noise parameters in the liquid neon are a factor of two to three larger
than those observed in argon. The probable cause for such a large amount of
noise is the poor gain characteristics of the PMTs and the use of an
additional amplifier physically far away from the PMT itself, in addition to
the factor of four longer integration period required for liquid neon scintillation. The solid line in Fig.~\ref{fig:PSDModel}
 shows the statistical model prediction in the ideal case where the additional
noise is set to zero. We note that the effect of the additional noise is most relevant for energies above 100 keVr.

To test the hypothesis that the poor PMT response negatively affected the PSD results, the signals in
each PMT were analyzed separately.
As expected given the single photoelectron
distributions, the bottom PMT provided as much as an order of magnitude better PSD compared to the top PMT above 150 keVr, dropping to a factor of about 2 at lower energies, although the combination of the two PMTs still provides an
improvement over just the bottom PMT alone. To achieve better overall PSD, more effort will be needed to
improve the response of the PMTs at liquid neon temperatures; as the two PMTs
exhibited different behaviors when cold, all PMTs intended for use in a liquid neon
detector may need to be cryogenically tested.  Another potential area of
improvement would be a more sophisticated discrimination statistic. Given the
complicated nature of scintillation in liquid neon (see
Sec.~\ref{sec:Temp}), a PSD analysis using more of the timing information
along the lines of the multi-bin method described in~\cite{Lippincott:2008} might
prove rewarding.


\section{Nuclear recoil scintillation efficiency}
\label{sec:LEff}
As mentioned in the introduction, it is well known that nuclear recoils produce less scintillation light in liquid noble gases than electronic recoils do. 
There are several processes
related to the linear energy transfer (LET) of the incoming radiation that are known to contribute to the lower scintillation
efficiency of nuclear recoils. The first of these is  the Lindhard
effect~\cite{Lindhard:1963}, where some of the energy deposited goes into
heat instead of the creation of molecules. The Lindhard effect
has been shown to be sufficient to explain quenching in
germanium~\cite{Benoit:2007}, but it is inadequate to explain \leff\, for the noble
gases. A second mechanism is known as bi-excitonic or Hitachi quenching~\cite{Hitachi:2005},
whereby
\begin{equation}
\mathrm{Ne}^* + \mathrm{Ne}^* \rightarrow \mathrm{Ne} + \mathrm{Ne}^+ + e^-.
\end{equation} 
When such a reaction occurs, the potential for creating two photons from the
two original exciton states has been reduced to one photon via recombination of the
single resulting ion. This reaction
is more likely to occur in high density tracks like those produced by nuclear
recoils. Recently, a model has been proposed to phenomenologically tie this
quenching to the stopping power~\cite{Mei:2008,Manzur:2010}. 
Lastly, some fraction of ionized electrons never recombine; if these ``escape
electrons'' are more likely for nuclear recoils than electronic recoils, this
process would contribute to a smaller \leff. This effect has been observed to lower \leff\, in
 liquid
xenon~\cite{Doke:2002,Manzur:2010,Sorensen:2011}.

The experimental setup for measuring the nuclear recoil scintillation
efficiency has been described in~\cite{Gastler:2009} in the context of a
measurement in liquid argon. One salient difference between the
current analysis and the argon analysis is the absence of MC simulations of
the neutron scattering. There are few experimental data on neon-neutron scattering, with Geant4 switching discretely from data to a model for neutrons of energy $<20$ MeV. There are efforts underway to improve this situation for future studies~\cite{MacMullin:2010}. 

Instead of fitting MC spectra to the data to determine the scintillation
efficiency, a simple Gaussian function is fit to the data to determine the mean of the observed
peak in units of keVee. Given the non-Gaussian nature of the single photoelectron distribution, we expect some skewing of the observed spectra towards high energies. In a simple MC of a similar situation in a larger detector, the contribution of this skew is $\approx5\%$ for the lowest energy point, below the level of our dominant uncertainties. Hence, our scientific judgment is that the uncertainty introduced by this skew does not contribute significantly to our final uncertainty estimate. The final expression for \leff\, is the
observed mean of the signal yield divided by the expected energy of the recoil as
predicted by Eq.~\ref{eq:NeutronScattering}.

\begin{figure*}[htbp]
\begin{center}
  \includegraphics*[width=\textwidth]{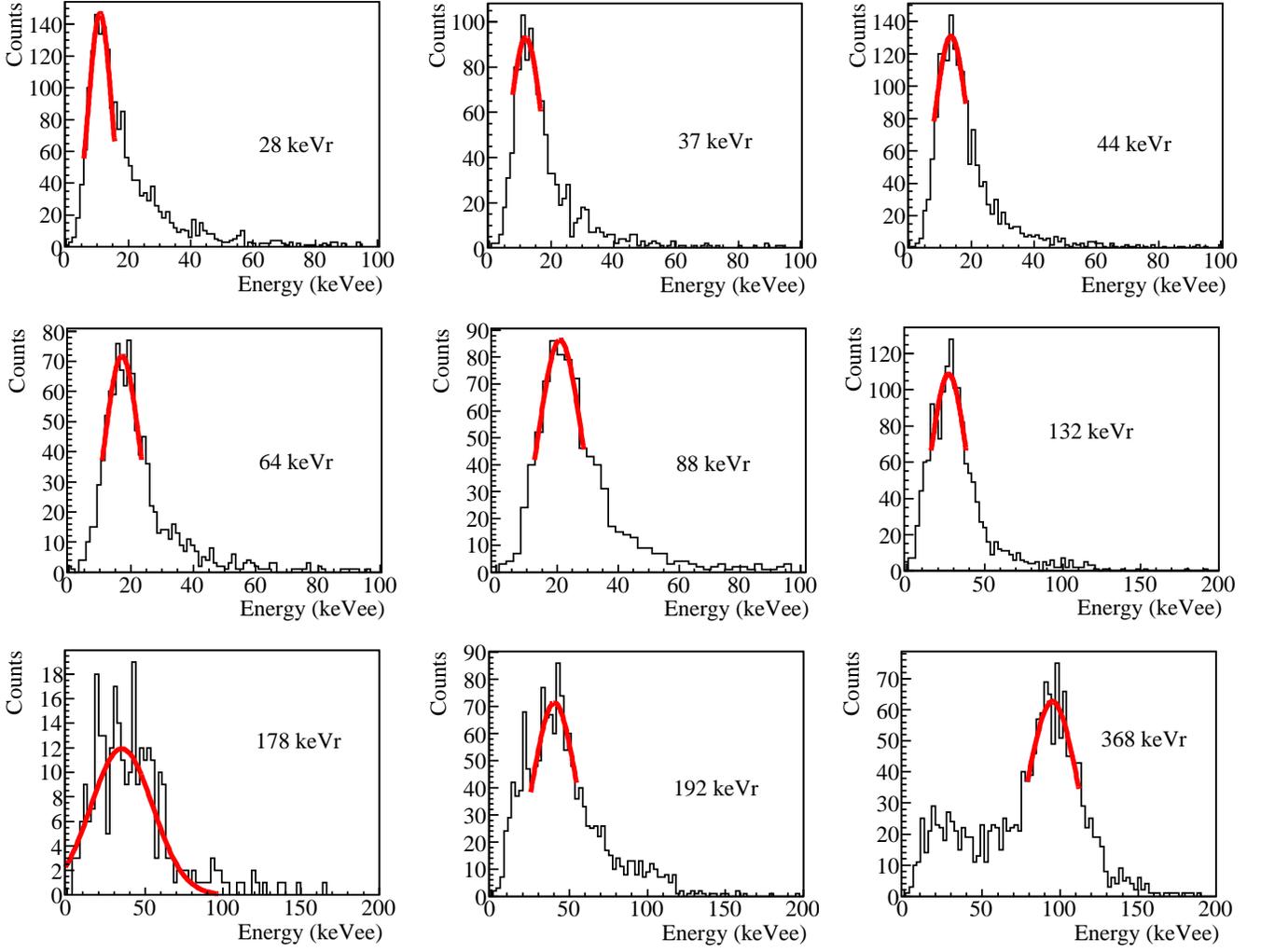}
\end{center}
\caption[Fits for \leff\, in neon]{(Color online) Fits for the nuclear recoil scintillation
  efficiency in liquid neon for all energies. All data sets include results from each
temperature setting, except for the data at 178 keVr that were only taken at
26.7 K.}
\label{fig:Peaks}
\end{figure*}

\begin{figure}[htbp]
\begin{center}
  \includegraphics[width=80mm]{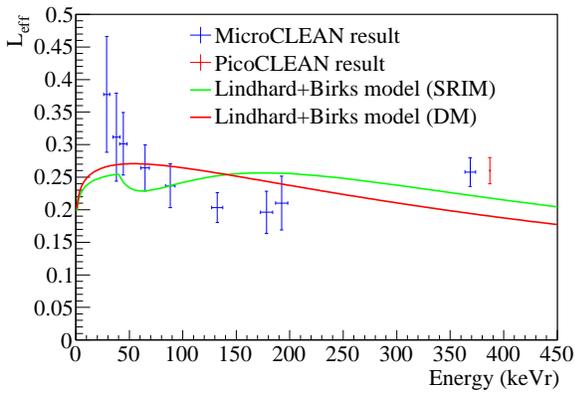}
\end{center}
\caption[Fits for \leff\, in neon]{(Color online) The observed
  nuclear recoil scintillation efficiency versus nuclear recoil energy in
  neon, along with the Lindhard+Birks model described in the text.}
\label{fig:neonLeff}
\end{figure}

Data were taken at nine different scattering angles corresponding to energies
ranging from $28.9$ keVr to $368.7$ keVr. Each point, with the exception of the point at 178
keVr, was taken at the three different operating temperatures. The results at each temperature are consistent with each other to
within the uncertainties of the measurement, so for the final determination of
\leff, all the data from each temperature are combined into one data
set. Figure~\ref{fig:Peaks} shows the final Gaussian fits at each energy, and
Figure~\ref{fig:neonLeff} shows the observed nuclear recoil scintillation
efficiency, along with the result from the previous measurement at 387
keVr~\cite{Nikkel:2007}. To obtain \leff\, we assume that the signal yield from electronic recoils scales linearly with energy (consistent with our observations between 40 keVee and 511 keVee) and compare to the corresponding signal yield at each nuclear recoil energy. The mean value of \leff\, thus obtained is 0.24. 

Because we expect there to be some contamination from multiple-scattering and gamma backgrounds into the distribution, as observed in argon data~\cite{Gastler:2009}, we limit the fit range in the region of the peak in the distribution; the fit range is iteratively chosen to encompass
plus and minus one sigma from the central value representing our estimate of the mean energy of the peak. The uncertainty associated with this
choice of fit range will be discussed in Sec.~\ref{sec:NeonLeffErrors}.  The average $\chi^2/$NDF for the fits shown in Fig.~\ref{fig:Peaks} is 1.07; the largest value of $\chi^2/$NDF is 2.5 for the 132.2 keVr peak, with all others below 1.3.

\subsection{Data selection cuts}
\begin{figure}[htbp]
\begin{center}
  \includegraphics[width=80mm]{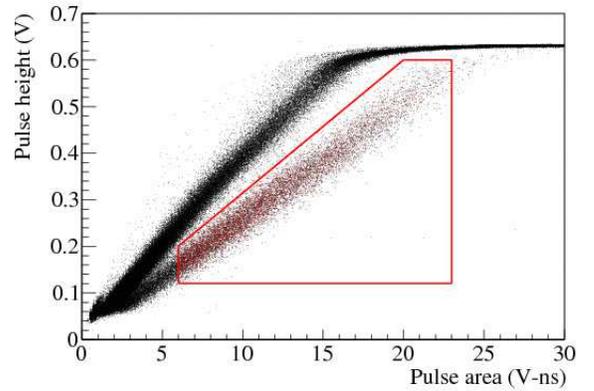}
\end{center}
\caption[PSD in the organic scintillator for neon runs]{(Color online) The PSD cut in the
  organic scintillator to select neutron events. Only events
  within the pentagon are included in the data set.}
\label{fig:OSCPSD}
\end{figure}

 One data selection cut requires that the two PMTs trigger within
40 ns of each other and that the asymmetry parameter is within
one sigma of the mean asymmetry value for the run. To select neutron-like events,
a PSD cut was applied to the organic scintillator data as shown in Fig.~\ref{fig:OSCPSD}. We also apply a TOF cut
between an event in the neon and the tagging event in the organic
scintillator. A final cut is applied based on the PSD described in the
previous section; this is a very loose cut, eliminating events farther than two
sigma from the measured mean value of  $\hat{f_p}$, for nuclear recoils at 26.7 K. 

\begin{figure}[htbp]
\begin{center}
  \includegraphics[width=80mm]{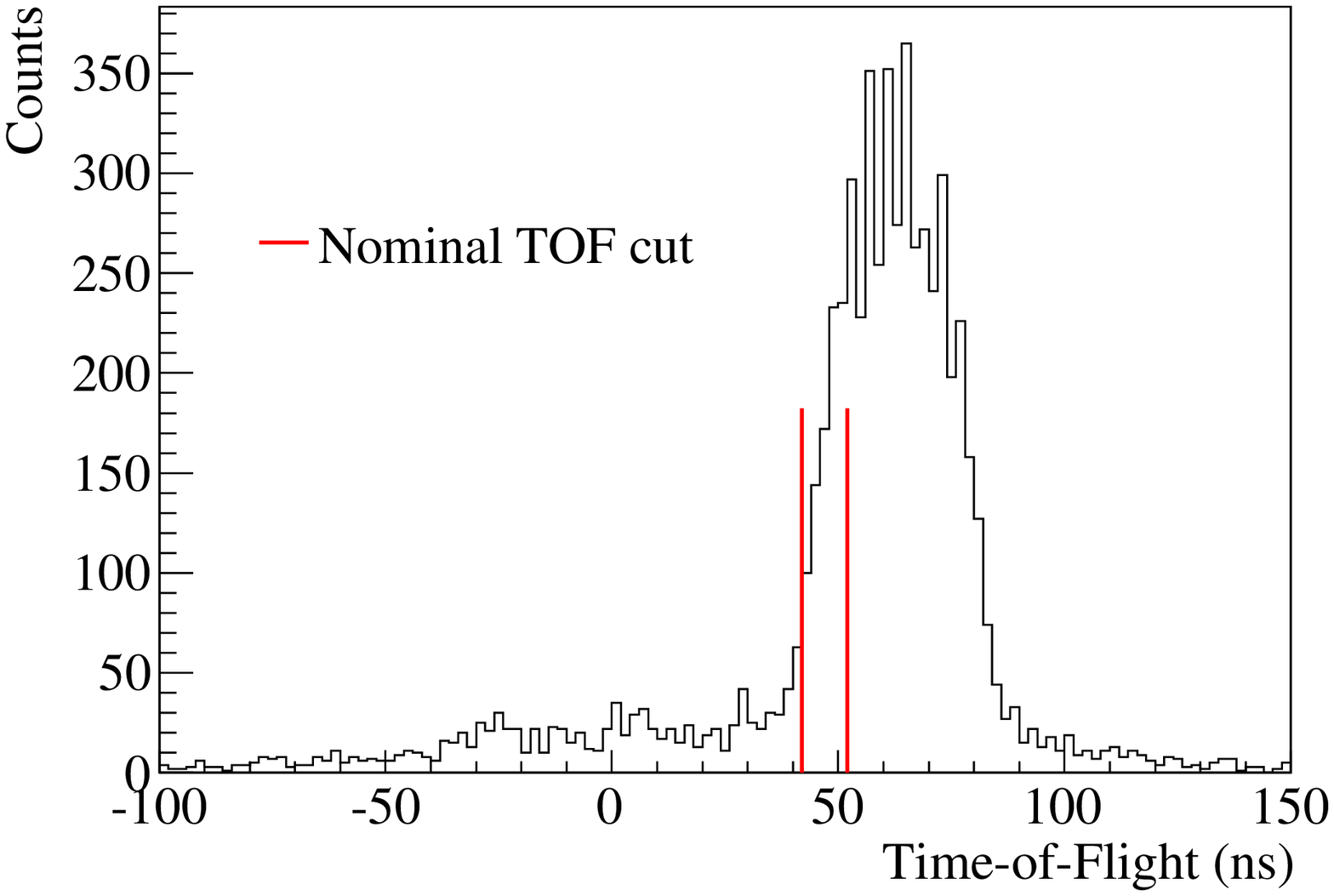}
\end{center}
\caption[Example TOF cut for 368.7 keV events]{The TOF cut window was 10 ns, and
  only events at the beginnning of the TOF peak were selected, as these events
  are more likely to be single-scatter events.}
\label{fig:TOFWindow}
\end{figure}

The TOF cut appears to be quite effective in selecting single-scatter nuclear recoil
events in liquid neon. For all runs, neutron events
formed a clear peak in the TOF spectrum. As single-scattering neutrons are
more likely to be found at the very beginning of the TOF peak, the TOF cut was
located at the front side of the peak. First, the total number of counts in the neutron peak was
found, and then the TOF distribution was scanned until the number of counts
in a 4-ns-wide bin exceeded $3\%$ of the total integral of the neutron
peak; this point set the left boundary of the TOF window. The width of the TOF acceptance window was chosen to be 10 ns. An
example of the TOF cut is shown in Fig.~\ref{fig:TOFWindow} for 368.7 keV events.

\begin{figure*}[htbp]
\begin{center}
  \includegraphics[width=0.7\textwidth]{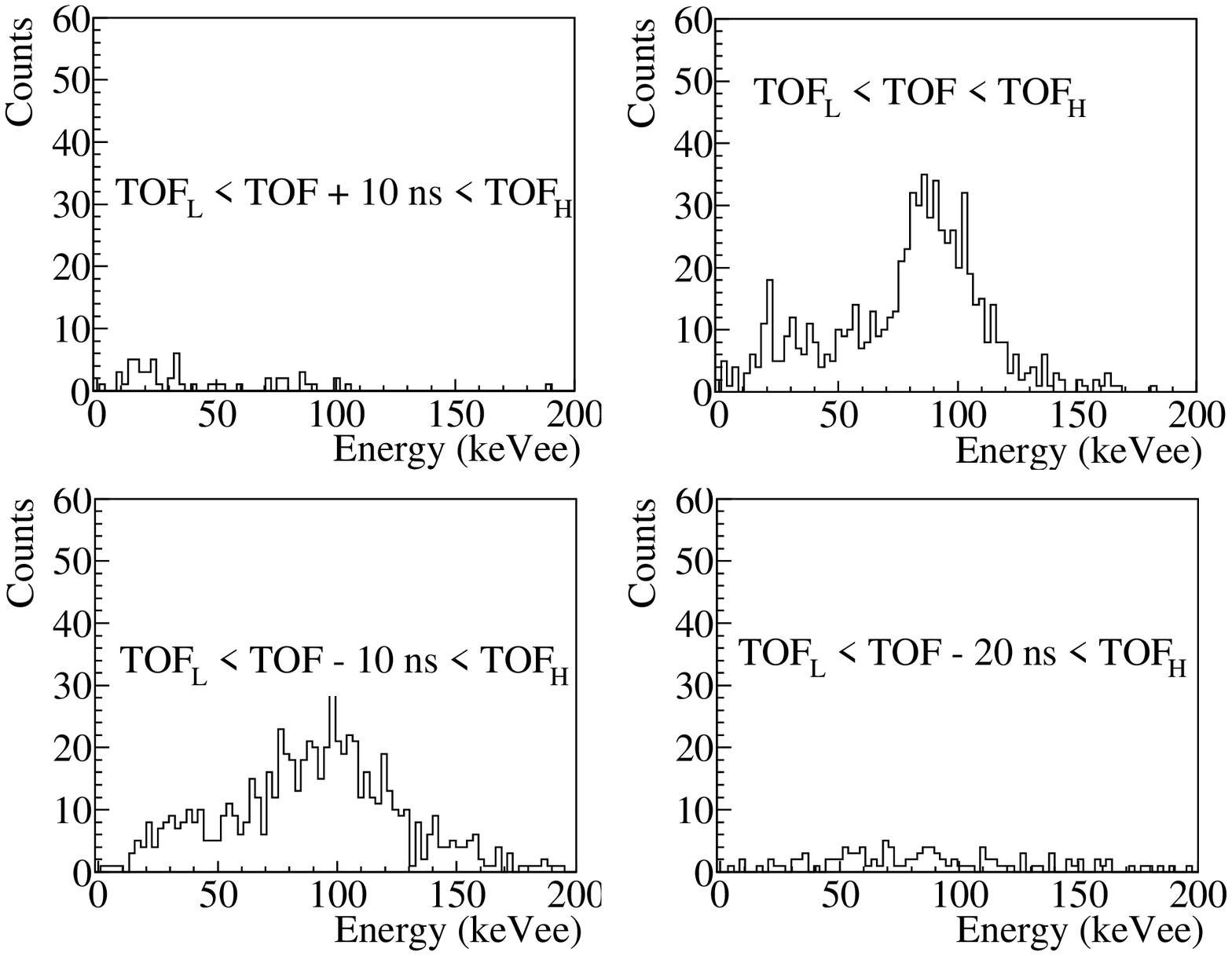}
\end{center}
\caption[The observed energy spectrum for 368.7 keV events versus TOF cut position]{The observed energy spectrum for 368.7 keVr events as the location of
  the TOF cut (as defined by $\mathrm{TOF_L}$ and $\mathrm{TOF_H}$) is
moved by $\pm 10$ ns and $+20$ ns. The nominal cut around the front side of
the TOF peak is the case where
$\mathrm{TOF_L} < \mathrm{TOF} < \mathrm{TOF_H}$. The degradation in the
observed peak as the TOF cut moves away from standard location  is clear.}
\label{fig:TOFHE}
\end{figure*}

\begin{figure*}[htbp]
\begin{center}
  \includegraphics[width=0.7\textwidth]{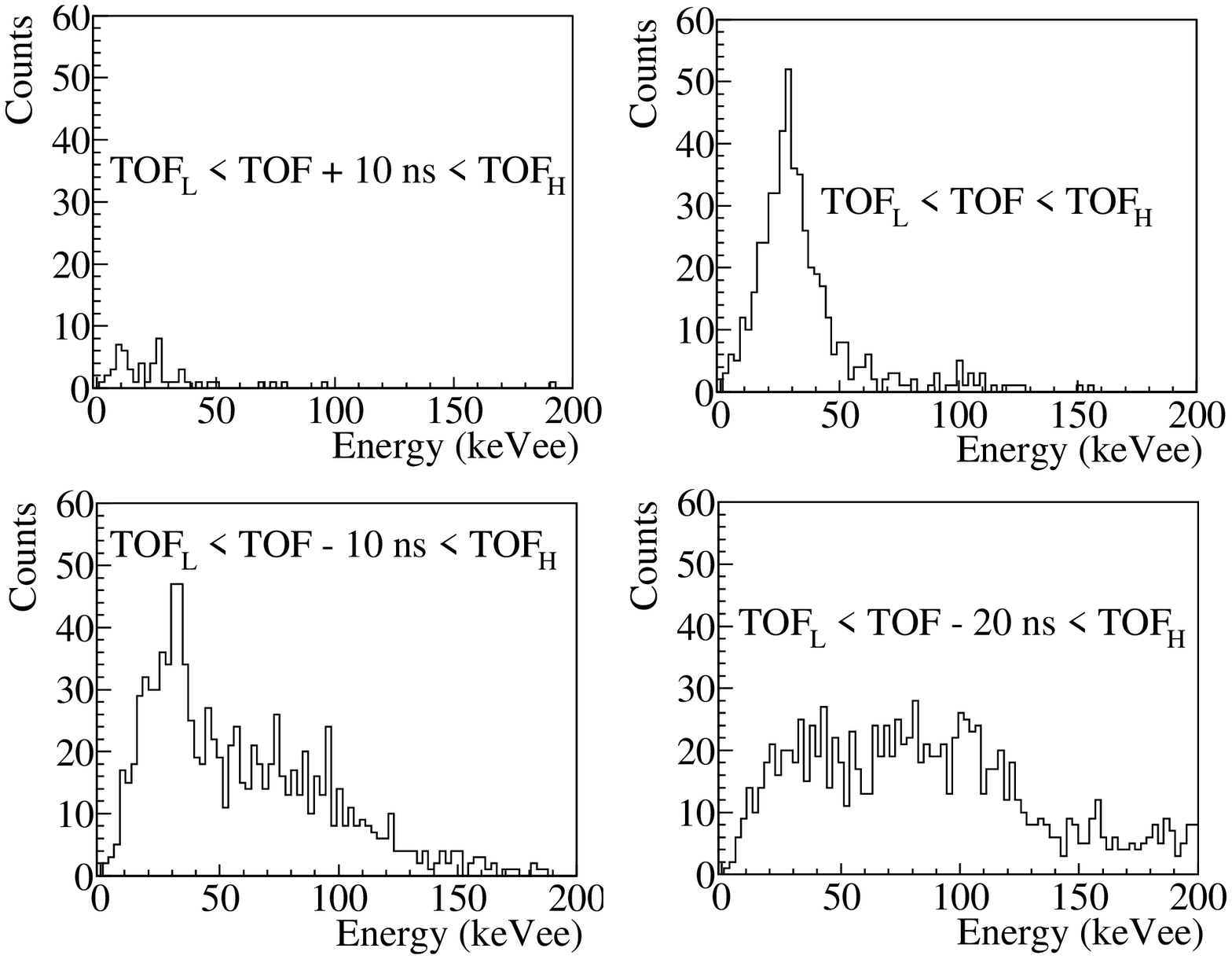}
\end{center}
\caption[The observed energy spectrum for 28.9 keV events versus TOF cut position]{The observed energy spectrum for 28.9 keVr events as the location of
  the TOF cut (as defined by $\mathrm{TOF_L}$ and $\mathrm{TOF_H}$) is
moved by $\pm 10$ ns and $+20$ ns. The nominal cut around the front side of
the TOF peak is the case where
$\mathrm{TOF_L} < \mathrm{TOF} < \mathrm{TOF_H}$. The degradation in the
observed peak as the TOF cut moves away from standard location  is clear.}
\label{fig:TOFLE}
\end{figure*}

Figures~\ref{fig:TOFHE}~and~\ref{fig:TOFLE} show the change in the observed
energy spectra produced by sliding the
location of the TOF cut window to earlier and later times for both the lowest
and highest energy runs. If the window is moved 10 ns earlier, very few events
pass the cut and no peak in energy is visible. At the standard cut location,
clear peaks in the energy spectra for both the low energy and high energy
points are visible. As the window moves to accept events with longer
times-of-flight, the energy peak loses definition due to the inclusion of
multiple-scattering backgrounds, becoming an almost flat
background when the TOF cut is moved to the center of the TOF peak (at 20 ns
beyond the standard cut). These results provide confidence that the standard
TOF cut at the front side of the TOF peak is
selecting single-scattering events and eliminating multiple-scattering backgrounds.  

\subsection{Error analysis}\label{sec:NeonLeffErrors}
There are a number of sources of uncertainty in this
measurement of \leff\, relating to both experimental parameters as well as data
selection cuts. There is a general uncertainty of $3\%$ from the energy calibration described above. Studies of the analysis cuts were performed independently to assess the
size of the systematic error associated with each. In each case, we took the difference between the value of \leff\, determined by the standard analysis and that determined with the changed cut value as our estimate of the systematic error associated with that cut. The size of the coincidence
window between the trigger times of the two PMTs was narrowed to 20 ns,
and the cut was dropped entirely to study its effect. Both the individual trigger cuts and the asymmetry cut were expanded to encompass
the region within two sigma of the mean for the entire run. The range in pulse area of the
organic scintillator PSD cut was first narrowed and then scanned across the entire
distribution of nuclear recoils (the lower band in Fig.~\ref{fig:OSCPSD}). To assess the uncertainty associated with the TOF cut, the width of
the TOF window was both narrowed to 6 ns and expanded to 16 ns, without
changing the location; as already discussed, moving the TOF cut window away
from the front of the TOF peak eliminated any observable peak, so the location of the cut was not included as a systematic uncertainty. Finally,
the analysis was repeated with no PSD cut applied to the neon data. To estimate the size of the uncertainty associated with the fit range in
Fig.~\ref{fig:Peaks}, the range was expanded to include the region within two
sigma of the central value. The uncertainty associated with the electron equivalent
energy calibration was determined to be 2$\%$. A source of uncorrelated error
is uncertainty in the location of the organic scintillator affecting the
predicted value of the nuclear recoil energy. An
uncertainty of 0.64 cm in the location of the organic scintillator around the detector corresponds to an
angular uncertainty of 1.3$^\circ$ at each position. A final source of uncertainty is the
trigger efficiency. Data for the three lowest energy positions were taken at
different hardware trigger thresholds such that the difference between the lowest and
highest threshold was larger than the average height of a single photoelectron
in each PMT. 

The statistical uncertainties in the fit results of Fig.~\ref{fig:Peaks} were
generally below the level of the other uncertainties in the measurement. The main source of uncertainty for the lower energies found in the experiment was
the choice of fit range, producing an uncertainty
of as large as $15\%$ in the 28.9 keVr data set. In all cases, the effect of
increasing the fit range was to increase the observed value of \leff. The results from changing the
hardware threshold were consistent with each other at each of the lowest three
energy points. At the highest energies, the
dominant source of uncertainty was the width of the TOF window. All sources of
uncertainty
were assumed to be independent and combined in quadrature to produce the final
uncertainty values shown in Fig.~\ref{fig:neonLeff}. Table~\ref{table:LEFF} lists the central values and the uncertainties associated with each identified source of error.

\begin{table*}[!ht]
\centering
\begin{tabular}{cc|cccccccccccc|c|c}\hline\hline
 & & \multicolumn{11}{c}{Systematic uncertainty} & & \\ \hline
Energy & \leff & $\Delta E$ & Energy & Fit & Relative & Absolute & Hardware & OSc & Asym. & \fP & TOF & Temp. & Total &  Stat. & Total \\
(keV)& & (keV) & calib. & range & trigger & trigger & threshold & cut & & cut &  size & ($\sigma$) & sys.  & uncert. & uncert. \\\hline
29 & 0.38 & 2 & 0.01 & 0.06 & 0.01 & 0.01 & 0.03 & 0 &0.01 & 0.01 & 0.02 & 0.02 & 0.08 & 0.04 & 0.09 \\
38 & 0.31 & 3 & 0.01& 0.04 & 0.02 & 0.02 & 0.01 & 0.02 & 0.01 & 0.02 & 0.02 & 0.01 & 0.06 & 0.03 & 0.07 \\
44 & 0.30 & 3 & 0.01& 0.02 & 0.01 & 0.01 & 0.02 & 0.01 & 0.01 & 0 & 0.01 & 0.02 & 0.04 & 0.03 & 0.05 \\
65 & 0.26 & 4 & 0.01 & 0.01 & 0.02 & 0.01 & N/A & 0.02 & 0 & 0.01 & 0.02 & 0.01 & 0.03 & 0.02 & 0.04 \\
88 & 0.24 & 4 & 0.01 & 0.02 & 0.01 & 0 & N/A & 0 & 0 & 0 & 0.02 & 0.01 & 0.03 & 0.01 & 0.03 \\
132 & 0.20 & 5 & 0.01 & 0 & 0.02 & 0 & N/A & 0 & 0 & 0 & 0 & 0.01 & 0.02 & 0.01 & 0.02 \\
178 & 0.20 & 5 & 0.01 & 0 & 0.01 & 0 & N/A & 0.01 & 0 & 0 & 0.02 & 0 & 0.03 & 0.01 & 0.03 \\
192 & 0.21 & 5 & 0.01 & 0 & 0.01 & 0 & N/A & 0 & 0.01 & 0 & 0.03 & 0.02 & 0.04 & 0.01 & 0.04 \\
369 & 0.26 & 5 & 0.01 & 0 & 0 & 0 & N/A & 0 & 0 & 0 & 0.01 & 0.02 & 0.02 & 0 & 0.02 \\
 \hline 
\end{tabular}
\caption[\leff\, parameters and errors]{The observed \leff\, parameters as a function of energy along with the estimated uncertainties for each of the sources of uncertainty listed in the text. The fit range column refers to expanding the fit range out to two sigma. The hardware threshold was varied for only the three lowest energy points. The ``OSc cut'' column refers to the location of the data cut used on the organic scintillator neutron detector. The ``TOF size'' refers to the size of the time-of-flight cut, which was both expanded to 16 ns and shortened to 6 ns, with both contributions included in that column.  The "Temp." column lists the standard deviation of the three temperature set points. All uncertainties that did not contribute at the hundredths level have been listed as zero.  }
\label{table:LEFF}
\end{table*}

\subsection{Discussion and the Lindhard+Birks Mode}

The first point to note about our results for \leff\, in liquid neon is an
upturn at low energies similar to one observed in our previous measurement of
\leff\, in liquid argon~\cite{Gastler:2009}, although the uncertainties on the lowest energy points are also the most significant. Both
hardware and software trigger effects were examined in attempts to explain
this upturn, with neither producing the observed result. As with the argon, we do not know whether this is a physically real effect. 

In the introduction, two main processes were listed as causes of the
lower scintillation efficiency of nuclear recoils in liquid noble gases. The
first was the Lindhard effect, where some energy goes into heat instead of the
creation of molecules. A second effect was Hitachi quenching, where two
excited atomic states interact to produce one ground state atom and one ion,
reducing the total number of Ne$_2$ molecules available to produce scintillation photons. Mei {\it et al.} have
developed a phenomenological model to account for these two
effects~\cite{Mei:2008}, and we apply that model here to the current results
drawing values for the stopping power from the tables of SRIM~\cite{Ziegler:1985} (``SRIM'' in Fig.
~\ref{fig:neonLeff}). We also use stopping powers obtained from Mei et al. from a second model for stopping power as described in~\cite{Mei:2008} (``DM'' in Fig.~\ref{fig:neonLeff}).

The result of fitting the Lindhard+Birks model to our data is shown in
Fig.~\ref{fig:neonLeff}. The Birks parameters, $kB$, for the SRIM and DM stopping powers are determined to be $1.99 \times 10^{-3}$ and $1.73 \times 10^{-3}$ MeV$^{-1}$ g cm$^{-2}$, respectively. This can be compared to the value of $1.12 \times 10^{-3}$ MeV$^{-1}$ g cm$^{-2}$ determined in~\cite{Mei:2008} from the single data point marked as ``PicoCLEAN result'' in Fig.~\ref{fig:neonLeff}.   The
kink that appears around 40 keVr in the SRIM curve is potentially mirrored in the upturn at
low energies but is not present in the DM model. In developing the Lindhard+Birks model for argon, Mei {\it
  et al.} found a similar kink from
the SRIM data that was not present in alternative
models~\cite{Mei:2008}.


\section{Scintillation time dependence}\label{sec:Temp}
The established model for scintillation in liquefied noble gases consists of two
exponentially decaying components representing the contribution from singlet and triplet
molecules. In our measurements of scintillation in liquid neon, although fast
and slow exponential behavior is observed, neither a two-component nor a three-component model is
sufficient to describe the time dependence. In addition, the observed time constants and intensities of
each component vary significantly with the temperature and pressure of the
liquid. In an attempt to understand the various components of liquid neon
scintillation and the temperature dependence, we examine average traces
from electronic recoil events with energies between 300 keVee and 400 keVee
for nine different pressure and temperature settings.  All data are derived from tagged $^{22}$Na events;
the trigger times of each event in the data sets are aligned and the events are summed together
to produce the average traces.

\begin{figure}[htbp]
\begin{center}
  \includegraphics[width=80mm]{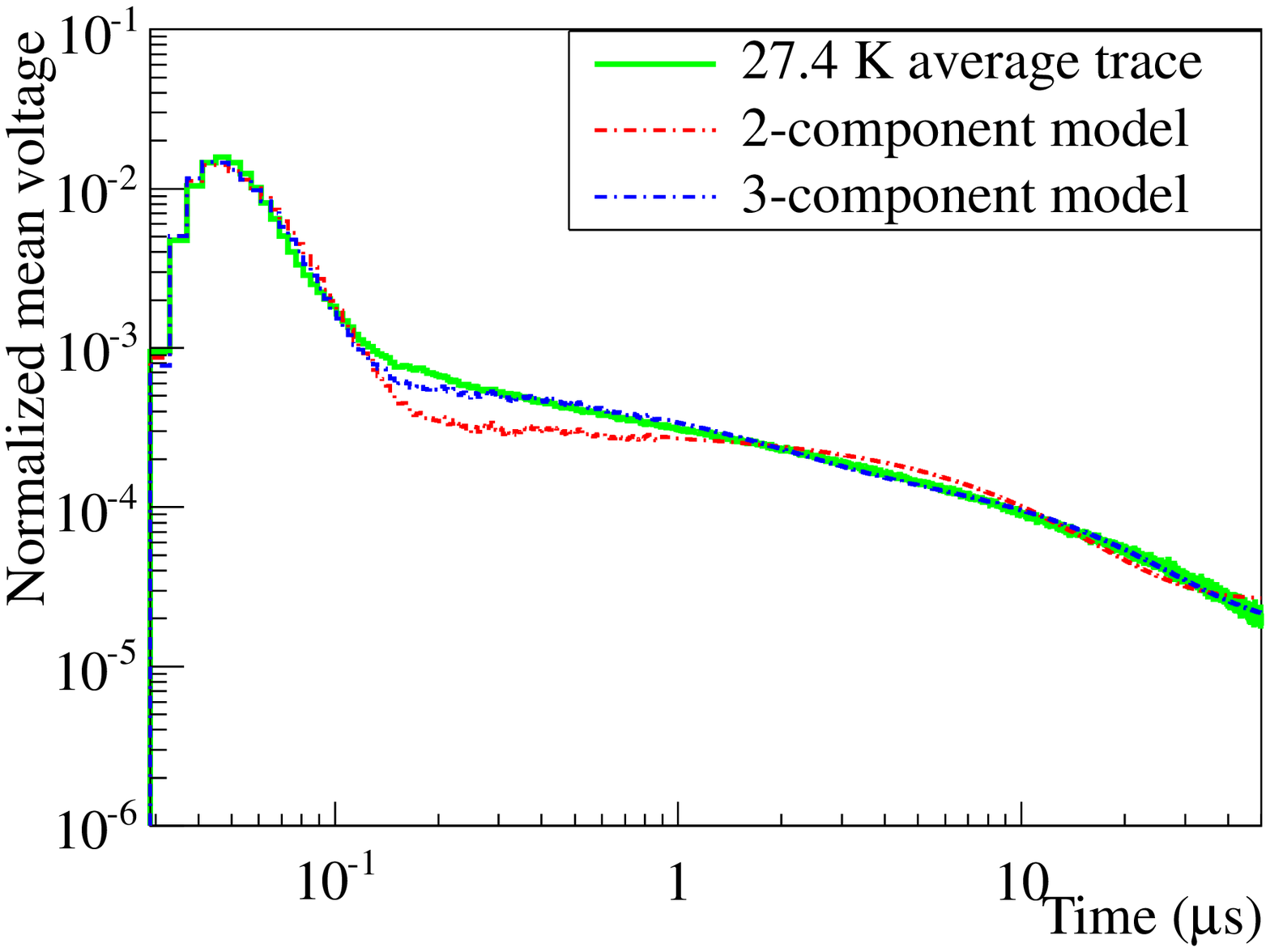}
\end{center}
\caption[Two- and three-exponential model for scintillation in liquid
neon]{(Color online) The average
  traces for 27.4 K data, along with fits to
  two-component and three-component exponential models. Neither model
  accurately describes the intermediate time regime.}
\label{fig:Neon3Exp}
\end{figure}

Figure~\ref{fig:Neon3Exp} shows an example trace for data taken at 27.4 K, 
along with fits to a two- and three-exponential
model. While the three-exponential model improves on the quality of
the overall fit, it still does not accurately model the data between
about 50 ns and 1 $\mu$s. We therefore adopt a four-exponential
mixture model~\cite{McLachlan:2000}. The exponential decay constants are denoted $\tau_i$,
and the relative weights,  which are non-negative and sum to 1, are denoted $I_i$. The mixture model for
the probability density function of the creation times of scintillation photons created by an energy
deposition at $t=0$ is
\begin{equation}
f(t) = \sum_{i=1}^4 \frac{I_i}{\tau_i} e^{-t/\tau_i}.
\end{equation} 
We model the expected
voltage trace produced by the PMTs as a convolution of an
experimentally determined impulse response function of the PMT, $h(t)$, and the
PDF model:
\begin{equation}
\left< V(t) \right> \propto \int_{t_0}^{t} h(t-s)f(s-t_0)ds,
\label{eq:model}
\end{equation}
where $t_0$ is the time of the energy deposit. We perform the fit by
minimizing the Matusita distance\cite{Matusita:1954} between the model and the data. 

\begin{figure}[htbp]
\begin{center}
  \includegraphics[width=90mm]{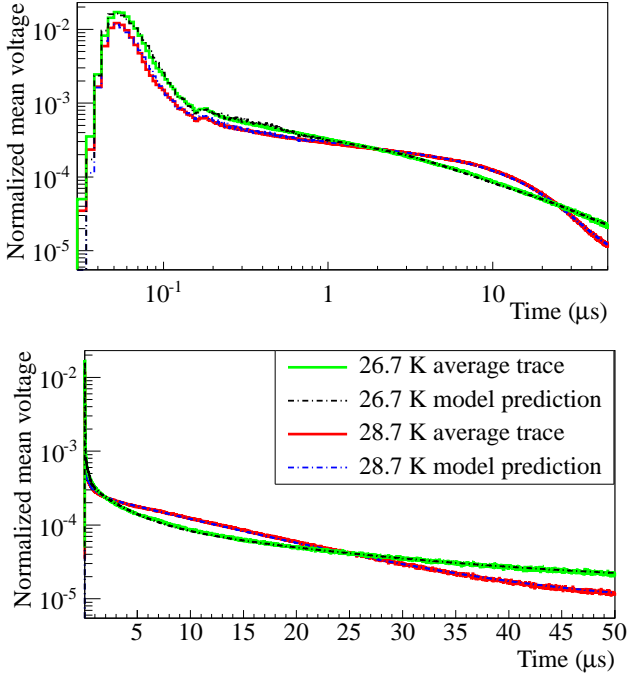}
\end{center}
\caption[Example fits to the four-component exponential]{(Color online) Examples of the four-component model fit to 26.7 K and 28.8 K data.}
\label{fig:KevinExample}
\end{figure}

\begin{figure}[htbp]
\begin{center}
  \includegraphics[width=80mm]{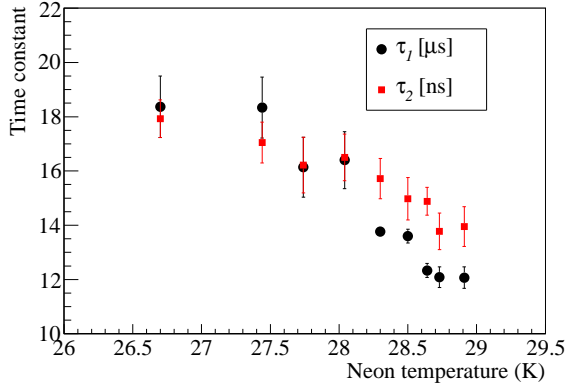}
\end{center}
\caption[Long and short time constants versus temperature]{(Color online) The fitted longest and shortest time constants, $\tau_1$ and $\tau_2$, for neon scintillation as a function of neon temperature. The error bars represent the combined estimated statistical and systematic uncertainties, derived as described in the text. }
\label{fig:tau12}
\end{figure}

\begin{figure}[htbp]
\begin{center}
  \includegraphics[width=80mm]{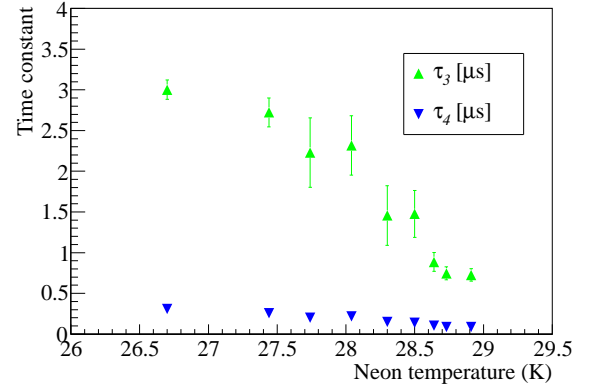}
\end{center}
\caption[Intermediate time constants versus temperature]{(Color online) The fitted intermediate time constants, $\tau_3$ and $\tau_4$, for neon scintillation as a function of neon temperature. The error bars represent the combined estimated statistical and systematic uncertainties, derived as described in the text. }
\label{fig:tau34}
\end{figure}

\begin{figure}[htbp]
\begin{center}
  \includegraphics[width=80mm]{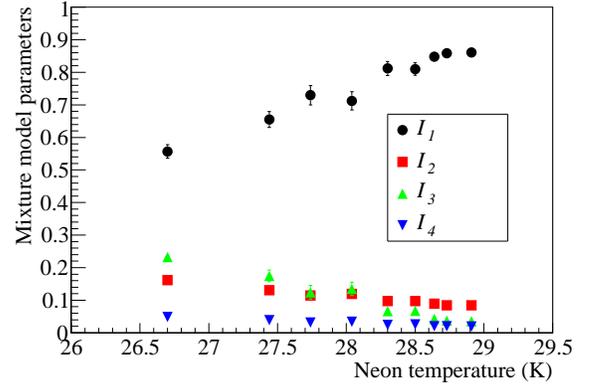}
\end{center}
\caption[Weight of the various components versus temperature]{(Color online) The fitted weights of the four-component model for neon scintillation as a function of neon temperature. The error bars represent the combined estimated statistical and systematic uncertainties, derived as described in the text. }
\label{fig:Ps}
\end{figure}

 Example fits to the 26.7 K
and 28.8 K data are shown in Fig.~\ref{fig:KevinExample}. We associate $\tau_1$
and $\tau_2$ with the longest and shortest time
components of the scintillation, respectively, and they are shown in
Fig.~\ref{fig:tau12} as a function of neon temperature. The short time
constant, $\tau_2$, should not be strictly associated with the decay of the
singlet state, as the fast component is also affected by the timing
characteristics of the TPB wavelength shifter and the
PMTs.  The remaining scintillation time constants,
representing intermediate time scales, are shown in
Fig.~\ref{fig:tau34}. The weights of each component as a function
of temperature are shown in Fig.~\ref{fig:Ps}, and all fit parameters
are reported in Table~\ref{table:NeonParameters}.

We estimate one sigma random uncertainties using a bootstrap resampling
scheme~\cite{Efron:1993}. We determine approximate estimates of the
systematic uncertainties associated with choice of fit window by
varying the end time of the integral in Eq.~\ref{eq:model}
between 44 and 52 $\mu$s. We estimate the systematic uncertainty
associated with our measurement of the single photoelectron response
of the PMTs by using several different estimates of $h(t)$ taken from
different data sets throughout the run. We also perform the fit for each PMT individually to estimate the uncertainties associated with the different PMTs. In
Figs.~\ref{fig:tau12}-\ref{fig:Ps}, the error bars represent the
combination in quadrature of the statistical and systematic
uncertainties. In many cases, the error bars are smaller than the size
of the marker. As a further consistency
check, the
analysis was repeated for events with energy between 50 and 150 keVee, and
all the trends discussed below were observed.

\begin{figure}[htbp]
\begin{center}
  \includegraphics[width=80mm]{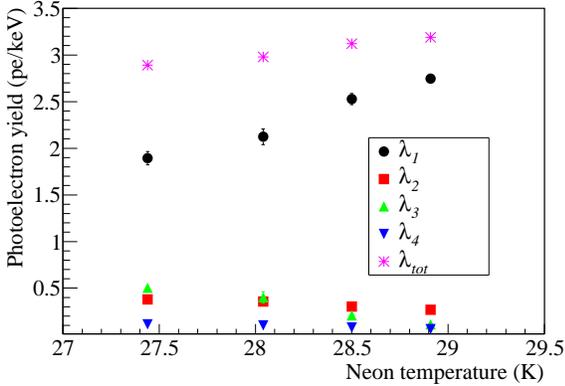}
\end{center}
\caption[Absolute scintillation component intensities  as a function of
temperature]{(Color online) The total signal yield plotted with the
  intensities of the long, short, intermediate and residual components as a function
  of temperature multiplied by the total signal yield of four runs taken over the
  course of three days.}
\label{fig:AbsoluteIntensity}
\end{figure}

We make a few general observations regarding the data before
continuing to a discussion of the physics. First, the observed time constants
decrease with increasing temperature and
pressure (Figs.~\ref{fig:tau12}-\ref{fig:tau34}). Second, the relative intensity of the long-lived  component increases with
temperature, primarily at the expense of the intermediate components with a
small contribution taken from the short-lived part (Fig.~\ref{fig:Ps}). 
The total signal yield also changes with temperature. Figure~\ref{fig:AbsoluteIntensity} shows
the contribution to the total signal yield by each of the four exponential decay processes in the model ($\lambda_i = I_i \times \lambda_{tot}$ where $\lambda_{tot}$ is the total signal yield) at four
temperatures taken within three days to minimize the effect of changing detector
conditions; the total signal yield increases with increasing
temperature, suggesting that not all of the increase in $I_1$ is due to
the decreases in the other components.

\begin{table*}[!ht]
\centering
\begin{tabular}{cccccccccccccccccc} \hline\hline
T & p & $\tau_1$ & $\sigma_{\tau_1}$ & $\tau_2$ & $\sigma_{\tau_2}$ & $\tau_3$ & $\sigma_{\tau_3}$ & $\tau_4$ & $\sigma_{\tau_4}$ &
$I_1$ & $\sigma_{I_1}$ & $I_2$ & $\sigma_{I_2}$ & $I_3$ & $\sigma_{I_3}$ &  $I_4$ & $\sigma_{I_4}$ \\ 
(K) & (mbar) & \multicolumn{2}{c}{(\mus)}  & \multicolumn{2}{c}{(ns)}  & \multicolumn{2}{c}{(\mus)} & \multicolumn{2}{c}{(\mus)} &  & &  & & &  & \\\hline
\multicolumn{18}{c}{Electronic recoils}\\\hline
26.7 & 816 & 18.37 & 1.13 & 17.93 & 0.70 & 3.00 & 0.12 & 0.31 & 0.03 & 0.56 & 0.02 & 0.16 & 0.01 & 0.23 & 0.02 & 0.049 & 0.002\\
27.4 & 1048 & 18.34 & 1.12 & 17.05 & 0.75 & 2.72 & 0.18 & 0.26 & 0.03 & 0.66 & 0.02 & 0.13 & 0.01 & 0.17 & 0.02 & 0.039 & 0.003\\
27.7 & 1160 & 16.14 & 1.10 & 16.22 & 1.02 & 2.23 & 0.43 & 0.20 & 0.04 & 0.73 & 0.03 & 0.11 & 0.01 & 0.12 & 0.02 & 0.032 & 0.004\\
28.0 & 1281 & 16.40 & 1.05 & 16.50 & 0.86 & 2.32 & 0.36 & 0.22 & 0.04 & 0.71 & 0.03 & 0.12 & 0.01 & 0.13 & 0.02 & 0.034 & 0.004\\
28.3 & 1380 & 13.77 & 0.15 & 15.72 & 0.74 & 1.46 & 0.37 & 0.15 & 0.02 & 0.81 & 0.02 & 0.097 & 0.005 & 0.07 & 0.01 & 0.025 & 0.004\\
28.5 & 1467 & 13.60 & 0.25 & 14.97 & 0.78 & 1.48 & 0.29 & 0.14 & 0.02 & 0.81 & 0.02 & 0.10 & 0.01 & 0.07 & 0.01 & 0.025 & 0.003\\
28.6 & 1520 & 12.33 & 0.26 & 14.88 & 0.51 & 0.89 & 0.12 & 0.10 & 0.01 & 0.85 & 0.01 & 0.089 & 0.004 & 0.04 & 0.01 & 0.020 & 0.002\\
28.7 & 1575 & 12.09 & 0.38 & 13.77 & 0.67 & 0.75 & 0.08 & 0.09 & 0.01 & 0.86 & 0.01 & 0.08 & 0.01 & 0.037 & 0.004 & 0.020 & 0.002\\
28.9 & 1662 & 12.07 & 0.39 & 13.95 & 0.73 & 0.73 & 0.08 & 0.09 & 0.01 & 0.86 & 0.01 & 0.08 & 0.01 & 0.035 & 0.004 & 0.019 & 0.002\\
\hline
\multicolumn{18}{c}{Nuclear recoils}\\\hline
27.7 & 1160 & 13.80 & 0.75 &  18.38 & 0.88 & 2.29 & 0.34 & 0.31 & 0.05 & 0.43 & 0.01& 0.35 & 0.01 & 0.142 & 0.005 &  0.08 & 0.01 \\
28.3 & 1380 & 10.99 & 0.51 & 17.49 & 1.02 & 1.26 & 0.31 & 0.175 & 0.04 & 0.52  & 0.02 & 0.32 & 0.01 & 0.105 & 0.004 & 0.05 & 0.01 \\
 \hline\hline
\end{tabular}
\caption[Time dependence parameters for scintillation in liquid neon]{Estimated time dependence parameters of the four-exponential model for scintillation in liquid neon. The long and short components are $\tau_1$ and $\tau_2$ respectively, and the intermediate components are $\tau_3$ and $\tau_4$. $I_i$ represents the relative weight of each component.  }
\label{table:NeonParameters}
\end{table*}

There are two final observations that can be made from the data. First, the
long time constant is affected by the presence of impurities, as the long time constant at
constant temperature increased during the purification cycle described in
Sec.~\ref{sec:DetStab}. Second, we perform a similar analysis on untagged nuclear
recoil data at two different temperatures (unfortunately, the tagged nuclear recoil data are limited
by statistics) and present the results at the bottom of
Table~\ref{table:NeonParameters}. 
The
parameter trends between the two nuclear
recoil data sets are consistent with the electronic recoil data, although the
long time constant for nuclear recoil events appears to be $1-2$ \mus\, shorter than the
corresponding time constants for electronic recoils. 


\subsection{Time dependence discussion}
Besides the very prompt light produced by the decay of singlet states, there
are three additional, distinct scintillation components.
The triplet lifetime in liquid neon has previously been measured to be 15.4,
2.9 and
3.9~\mus~\cite{Nikkel:2007, Suemoto:1979, Michniak:2002}, with the
large discrepancy between the first and second two measurements attributed to
impurities in the liquid in the latter two. In solid neon, the long exponential time constant
has been measured to be 3.3 and 3.9~\mus~\cite{Suemoto:1979,Michniak:2002}. A
very slow tail in the solid has also been observed with a
lifetime of 560~\mus, attributed to excitations of the 3s$_{12}^3$P$_2$ atomic state~\cite{Packard:1970}; in one measurement of
the liquid, an excited atomic triplet state was observed to decay non-exponentially,
disappearing within several microseconds~\cite{Suemoto:1979}. Measurements in gaseous neon
suggest that long-lived $^1$P$_1$ and $^3$P$_1$ atomic states can react via
three-body collisions to form radiative
molecular states~\cite{Leichner:1973}; the lifetime of the long-lived molecular state in gas,
presumably equivalent to the triplet state, was measured to be
6.6~\mus~\cite{Oka:1974}. Lastly, as the temperature of the gas is reduced,
the emission due to molecular states disappears, leaving only the atomic
spectra behind~\cite{Packard:1970}. 

Before further discussing the current neon data, we
compare the results to the situation in argon and helium. First, the data presented in~\cite{Lippincott:2008} for 
scintillation in liquid argon were taken at many different temperatures
covering a larger absolute temperature range than that examined in liquid neon, and there was no change in the
observed triplet lifetime or signal yield in argon. 

In liquid helium, the triplet lifetime is 13 s~\cite{McKinsey:1999}. Two other characteristic time scales were
observed in scintillation in liquid helium. First, metastable He(2$^3$S) atoms
created by ionizing radiation have been observed to last for 8000 s in
vacuum~\cite{Keto:1974a,Woodworth:1975}; in liquid, these states only last for
15~\mus, and their disappearance is accompanied by the appearance of vibrationally
excited triplet molecules, implying the creation of molecules by collisions
of the excited atomic states~\cite{Koymen:1990}. Second, a $t^{-1}$ component
has been observed over a finite time window and attributed to collisions of long-lived triplet states
leading to the production of the shorter-lived singlet states which promptly
decay~\cite{McKinsey:2003}. King and Voltz developed a model to describe slow
emission light produced by scintillators via triplet-triplet collisions, neglecting the effects
of the radiative decay of triplet molecules~\cite{King:1966}. The KV model was adapted by McKinsey et al. to describe the  $t^{-1}$ component in helium over a finite time window, although the exact form of the power law depends on the assumed track geometry~\cite{McKinsey:2003}.

Finally, an important similarity between liquid helium and liquid neon that sets them
apart from the heavier noble gases is that free
electrons form bubbles or ``localized'' states in both
liquids~\cite{Bruschi:1972,Loveland:1972,Kuper:1961,Jortner:1965,Springett:1967}.
These bubble states occur
because the  electron-atom repulsion due to Pauli exclusion is strong
relative to the weak
polarization of helium and neon atoms, as evidenced by a positive scattering length for
low energy electrons in helium and neon~\cite{OMalley:1963,Springett:1968,Miyakawa:1969}. In the heavier noble gases, the
polarization term becomes stronger and overwhelms the repulsive
pseudopotential. While electron bubbles would influence recombination rates,
affecting the timing of scintillation on short time scales as well as the
number density of singlet and triplet molecules, it is unlikely that they
are important for the late time behavior. However, in liquid helium, He$_2^{*}$ molecules also form bubble
states~\cite{Dennis:1969,Hill:1971,Hickman:1971}. Similar studies and
calculations have not been
performed for liquid neon, but it is certainly possible that Ne$_2^{*}$ could
also form bubble states, affecting the long-lived triplet
molecules and providing another difference between
scintillation in liquid neon and liquid argon.

The comparison to liquid helium seems to provide the most illumination,
particularly as free electrons form bubbles in both liquids. As in the current
measurements, four time components have been
observed in liquid helium scintillation: a prompt decay from singlet
molecules, a slow exponential decay from triplet molecules, an intermediate exponential decay
from atomic states that can also react to
form  molecules, and a $t^{-1}$ component from triplet molecules reacting to
produce singlet molecules.

In neon, we likewise attribute the slowest and fastest ($\tau_1$ and $\tau_2$) components to singlet and
triplet molecules. The number of promptly created singlet molecules
decreases with increasing temperature and pressure, while the number of
triplet molecules greatly increases. We know that there is a connection
between the recoil track density and the production of singlet and triplet
molecules, and a possible contributor to the changing intensities of each component is the changing density of
excitations with temperature. From measurements of
the electron mobility and the Einstein relation, the diffusion constant for
localized electrons increases by $30\%$ between 26.7 and 28.8
K~\cite{Bruschi:1972}. Recalling that one channel for the
production of metastable molecules is ion-electron recombination, in a simplified picture one
expects a larger diffusion constant for electrons to result in a greater
maximal separation between ions and recombining electrons. This could produce
a greater proportion of triplet molecules given the necessity of an electron
spin flip, a mechanism that has been proposed to explain the behavior of xenon
scintillation under an electric field~\cite{Kwong:2010}.  However, the increase in $I_1$ with increasing temperature is much greater than the
corresponding 
decrease in $I_2$, requiring some additional explanation that may be found
in the intermediate components. 

Continuing to examine the triplet states, we must also conclude that the
apparent lifetime of the triplet molecules changes as a function of the temperature or pressure of the liquid
environment. Impurities in liquid noble gases can
shorten the apparent triplet lifetime via absorption or non-radiative
collisions. However, in the data of
Table~\ref{table:NeonParameters}, the shorter lifetime observed at higher
temperature and pressure is not accompanied by a decrease in $I_1$; in
fact, the contribution of the triplet component greatly increases with decreasing $\tau_1$,
contrary to what one would expect if decreasing $\tau_1$ was caused by impurities.

One possible mechanism that could explain the changing triplet lifetime
without an accompanying reduction in $I_1$ is the
presence of molecular bubble states analogous to those found in liquid helium;
one could easily imagine that the radiative
lifetime of the molecule changes depending on whether it resides within a
bubble or not. If Ne$_2^*$ molecules are weakly bound in
bubbles at these temperatures, then some fraction of molecules could be
localized in
bubbles with an additional fraction freely existing in the liquid, leading to the
following model for the decay rate of triplet molecules, $d\langle N_t\rangle/dt$:
\begin{equation}\label{eq:bubble}
\frac{d\langle N_t\rangle}{dt} = -f(p,T)\frac{<N_t>}{\tau_b} -
\left[1-f(p,T)\right]\frac{<N_t>}{\tau_{\bar{b}}}.
\end{equation}
Here, $f(p,T)$ is a function of temperature and pressure representing the fraction of
molecules contained in bubbles, $\tau_b$ is the lifetime of molecules in
bubbles and $\tau_{\bar{b}}$ is the lifetime of molecules outside of bubbles. Although such a model could account for the observed behavior, it is purely speculative given our lack of theoretical guidance for
whether neon molecules truly do form stable bubbles. 

There are two intermediate components, one with a lifetime of order $100$ ns and the other with a lifetime of order $1\,\mu$s. We can again draw a parallel to the situation in liquid helium and
attribute some of this component to collisions between triplet molecules seeding the creation
of singlet molecules that immediately decay as in the King and
Voltz theory. However, the KV theory cannot entirely account for the
intermediate exponential behavior, even if one modifies the KV model
 to include the decay of the triplet state. Therefore, we attribute the longer-lived
 intermediate exponential component to
either excited atomic states producing singlet molecules that immediately
decay or to the decay of the long-lived atomic states
themselves. In the first case, the observed lifetime $\tau_3$ is
equivalent to the reaction rate of  excited atomic
states with ground-state Ne atoms; if this collision rate increases with temperature or pressure (perhaps
because of increasing diffusion or bubble fluctuations), the observed lifetime
would decrease. In the latter case, $\tau_3$ would be some combination of the
true radiative lifetime of the atomic excitation and a quenching reaction. In
contrast to the very long-lived exponential channel, the decreasing $\tau_3$ \emph{is} accompanied by a
decreasing $I_3$, suggesting the existence of a non-radiative channel. Taking
the argument one step 
further, if we combine the two scenarios with the modification that three-body
collisions of excited atomic states create {\it triplet} molecules, the
non-radiative channel could result in the production of
triplet molecules and the observed increase in $I_1$. This interpretation could be
consistent with the disappearance of a triplet atomic spectrum over several
microseconds as observed by Suemoto and Kanzaki~\cite{Suemoto:1979}.

In the future, understanding the effects of an applied electric field on liquid neon
scintillation might illuminate some of the temperature/pressure dependencies
observed in this work. A second study that would be useful in assessing
whether molecules reside in bubble states would be further laser spectroscopy; in
liquid helium, molecular bubbles were first discovered in the absence of
wavelength shifts for rovibrational structure in emission spectra of He$_2^*$
between gaseous and liquid helium. These studies could be accompanied by
theoretical work on the potential existence of molecular bubbles in liquid
neon. Third, the experimental conditions of MicroCLEAN
required that the liquid follow the saturation line, rendering it impossible
to separate the temperature effects from those associated with pressure. It is likely that  bubble interactions depend on the pressure
more strongly than on the liquid temperature, and an experiment capable of probing
the liquid phase space away from the saturation line and at higher
temperatures could prove interesting.
\linebreak

\section{Conclusion}

In this paper, we have reported measurements of scintillation light. With PMTs immersed directly in liquid neon, we have observed a substantial signal yield of up to ($3.5 \pm 0.4$) photoelectrons/keVee. The observed signal yield is greater than what would be required for a $pp$-neutrino measurement in CLEAN and much larger than the signal yield assumed in previous simulations of a tonne-scale CLEAN detector~\cite{Coakley:2004, Boulay:2005}. We have demonstrated the use of activated charcoal as a purifier to remove light reducing impurities in the liquid. We have quantified the performance of the prompt ratio PSD method for liquid neon by measuring the ERC for a nuclear recoil acceptance probability of approximately 0.5 between 50 keVr and 300 keVr, although we believe our measurement to be an upper limit on the discrimination power achievable with liquid neon due to noise in our system. As in previous studies of liquid argon, we observe a convergence of the \fP\, parameter at low energies. We have also made a measurement of \leff\, in liquid neon. Finally, we have observed a very interesting time dependence of the scintillation of liquid neon, consisting of four distinct components and a clear temperature/pressure dependence in both intensity and timing. 

\begin{acknowledgments}
We gratefully acknowledge Dongming Mei for useful discussion and data on the stopping power of neon. We thank Joshua Klein, Stan Seibert, Franco Giuliani and Thomas Caldwell for useful discussion and for their work on the RAT simulation package.

This work was supported by a Packard Science and Engineering Fellowship, Yale University,  and the U.S. Department of Energy. Computer simulations were supported in part by the facilities and staff of the Yale University Faculty of Arts and Sciences High Performance Computing Center.
\end{acknowledgments}


\begin{thebibliography}{64}
\expandafter\ifx\csname natexlab\endcsname\relax\def\natexlab#1{#1}\fi
\expandafter\ifx\csname bibnamefont\endcsname\relax
  \def\bibnamefont#1{#1}\fi
\expandafter\ifx\csname bibfnamefont\endcsname\relax
  \def\bibfnamefont#1{#1}\fi
\expandafter\ifx\csname citenamefont\endcsname\relax
  \def\citenamefont#1{#1}\fi
\expandafter\ifx\csname url\endcsname\relax
  \def\url#1{\texttt{#1}}\fi
\expandafter\ifx\csname urlprefix\endcsname\relax\def\urlprefix{URL }\fi
\providecommand{\bibinfo}[2]{#2}
\providecommand{\eprint}[2][]{\url{#2}}

\bibitem[{\citenamefont{McKinsey and Doyle}(2000)}]{McKinsey:2000}
\bibinfo{author}{\bibfnamefont{D.~N.} \bibnamefont{McKinsey}} \bibnamefont{and}
  \bibinfo{author}{\bibfnamefont{J.~M.} \bibnamefont{Doyle}},
  \bibinfo{journal}{J. Low Temp. Phys.} \textbf{\bibinfo{volume}{118}},
  \bibinfo{pages}{153 } (\bibinfo{year}{2000}).

\bibitem[{\citenamefont{Boulay et~al.}(2005)\citenamefont{Boulay, Hime, and
  Lidgard}}]{Boulay:2005}
\bibinfo{author}{\bibfnamefont{M.~G.} \bibnamefont{Boulay}},
  \bibinfo{author}{\bibfnamefont{A.}~\bibnamefont{Hime}}, \bibnamefont{and}
  \bibinfo{author}{\bibfnamefont{J.}~\bibnamefont{Lidgard}},
  \bibinfo{journal}{Nucl. Phys. B, Proc. Suppl.}
  \textbf{\bibinfo{volume}{143}}, \bibinfo{pages}{486} (\bibinfo{year}{2005}).

\bibitem[{\citenamefont{McKinsey and Coakley}(2005)}]{McKinsey:2005b}
\bibinfo{author}{\bibfnamefont{D.~N.} \bibnamefont{McKinsey}} \bibnamefont{and}
  \bibinfo{author}{\bibfnamefont{K.~J.} \bibnamefont{Coakley}},
  \bibinfo{journal}{Astropart. Phys.} \textbf{\bibinfo{volume}{22}},
  \bibinfo{pages}{355 } (\bibinfo{year}{2005}).

\bibitem[{\citenamefont{McKinsey}(2007)}]{McKinsey:2007}
\bibinfo{author}{\bibfnamefont{D.}~\bibnamefont{McKinsey}},
  \bibinfo{journal}{Nucl. Phys. B, Proc. Suppl.}
  \textbf{\bibinfo{volume}{173}}, \bibinfo{pages}{152 } (\bibinfo{year}{2007}).

\bibitem[{\citenamefont{Jungman et~al.}(1996)\citenamefont{Jungman,
  Kamionkowski, and Griest}}]{Jungman:1996}
\bibinfo{author}{\bibfnamefont{G.}~\bibnamefont{Jungman}},
  \bibinfo{author}{\bibfnamefont{M.}~\bibnamefont{Kamionkowski}},
  \bibnamefont{and} \bibinfo{author}{\bibfnamefont{K.}~\bibnamefont{Griest}},
  \bibinfo{journal}{Phys. Rep.} \textbf{\bibinfo{volume}{267}},
  \bibinfo{pages}{195} (\bibinfo{year}{1996}).

\bibitem[{\citenamefont{Nikkel et~al.}(2008)\citenamefont{Nikkel, Hasty,
  Lippincott, and McKinsey}}]{Nikkel:2007}
\bibinfo{author}{\bibfnamefont{J.~A.} \bibnamefont{Nikkel}},
  \bibinfo{author}{\bibfnamefont{R.}~\bibnamefont{Hasty}},
  \bibinfo{author}{\bibfnamefont{W.~H.} \bibnamefont{Lippincott}},
  \bibnamefont{and} \bibinfo{author}{\bibfnamefont{D.~N.}
  \bibnamefont{McKinsey}}, \bibinfo{journal}{Astropart. Phys.}
  \textbf{\bibinfo{volume}{29}}, \bibinfo{pages}{161} (\bibinfo{year}{2008}).

\bibitem[{\citenamefont{Harrison et~al.}(2006)\citenamefont{Harrison,
  Lippincott, McKinsey, and Nikkel}}]{Harrison:2006}
\bibinfo{author}{\bibfnamefont{M.~K.} \bibnamefont{Harrison}},
  \bibinfo{author}{\bibfnamefont{W.~H.} \bibnamefont{Lippincott}},
  \bibinfo{author}{\bibfnamefont{D.~N.} \bibnamefont{McKinsey}},
  \bibnamefont{and} \bibinfo{author}{\bibfnamefont{J.~A.}
  \bibnamefont{Nikkel}}, \bibinfo{journal}{Nucl. Instrum. Meth. A}
  \textbf{\bibinfo{volume}{570}}, \bibinfo{pages}{556} (\bibinfo{year}{2006}).

\bibitem[{\citenamefont{Packard et~al.}(1970)\citenamefont{Packard, Reif, and
  Surko}}]{Packard:1970}
\bibinfo{author}{\bibfnamefont{R.~E.} \bibnamefont{Packard}},
  \bibinfo{author}{\bibfnamefont{F.}~\bibnamefont{Reif}}, \bibnamefont{and}
  \bibinfo{author}{\bibfnamefont{C.~M.} \bibnamefont{Surko}},
  \bibinfo{journal}{Phys. Rev. Lett.} \textbf{\bibinfo{volume}{25}},
  \bibinfo{pages}{1435} (\bibinfo{year}{1970}).

\bibitem[{\citenamefont{Lippincott et~al.}(2008)\citenamefont{Lippincott,
  Coakley, Gastler, Hime, Kearns, McKinsey, Nikkel, and
  Stonehill}}]{Lippincott:2008}
\bibinfo{author}{\bibfnamefont{W.~H.} \bibnamefont{Lippincott}},
  \bibinfo{author}{\bibfnamefont{K.~J.} \bibnamefont{Coakley}},
  \bibinfo{author}{\bibfnamefont{D.}~\bibnamefont{Gastler}},
  \bibinfo{author}{\bibfnamefont{A.}~\bibnamefont{Hime}},
  \bibinfo{author}{\bibfnamefont{E.}~\bibnamefont{Kearns}},
  \bibinfo{author}{\bibfnamefont{D.~N.} \bibnamefont{McKinsey}},
  \bibinfo{author}{\bibfnamefont{J.~A.} \bibnamefont{Nikkel}},
  \bibnamefont{and} \bibinfo{author}{\bibfnamefont{L.~C.}
  \bibnamefont{Stonehill}}, \bibinfo{journal}{Phys. Rev. C}
  \textbf{\bibinfo{volume}{78}}, \bibinfo{pages}{035801}
  (\bibinfo{year}{2008}).

\bibitem[{\citenamefont{Boulay et~al.}(2009)\citenamefont{Boulay, Cai, Chen,
  Golovko, Harvey, Mathew, Lidgard, McDonald, Pasuthip, Pollman
  et~al.}}]{Boulay:2009}
\bibinfo{author}{\bibfnamefont{M.~G.} \bibnamefont{Boulay}},
  \bibinfo{author}{\bibfnamefont{B.}~\bibnamefont{Cai}},
  \bibinfo{author}{\bibfnamefont{M.}~\bibnamefont{Chen}},
  \bibinfo{author}{\bibfnamefont{V.~V.} \bibnamefont{Golovko}},
  \bibinfo{author}{\bibfnamefont{P.}~\bibnamefont{Harvey}},
  \bibinfo{author}{\bibfnamefont{R.}~\bibnamefont{Mathew}},
  \bibinfo{author}{\bibfnamefont{J.~J.} \bibnamefont{Lidgard}},
  \bibinfo{author}{\bibfnamefont{A.~B.} \bibnamefont{McDonald}},
  \bibinfo{author}{\bibfnamefont{P.}~\bibnamefont{Pasuthip}},
  \bibinfo{author}{\bibfnamefont{T.}~\bibnamefont{Pollman}},
  \bibnamefont{et~al.}, \bibinfo{journal}{arxiv:0904.2930v1 [astro-ph.IM]}
  (\bibinfo{year}{2009}).

\bibitem[{\citenamefont{Benetti et~al.}(2008)\citenamefont{Benetti, Acciarri,
  Adamo, Baibussinov, Baldo-Ceolin, Belluco, Calaprice, Calligarich, Cambiaghi,
  Carbonara et~al.}}]{Benetti:2007}
\bibinfo{author}{\bibfnamefont{P.}~\bibnamefont{Benetti}},
  \bibinfo{author}{\bibfnamefont{R.}~\bibnamefont{Acciarri}},
  \bibinfo{author}{\bibfnamefont{F.}~\bibnamefont{Adamo}},
  \bibinfo{author}{\bibfnamefont{B.}~\bibnamefont{Baibussinov}},
  \bibinfo{author}{\bibfnamefont{M.}~\bibnamefont{Baldo-Ceolin}},
  \bibinfo{author}{\bibfnamefont{M.}~\bibnamefont{Belluco}},
  \bibinfo{author}{\bibfnamefont{F.}~\bibnamefont{Calaprice}},
  \bibinfo{author}{\bibfnamefont{E.}~\bibnamefont{Calligarich}},
  \bibinfo{author}{\bibfnamefont{M.}~\bibnamefont{Cambiaghi}},
  \bibinfo{author}{\bibfnamefont{F.}~\bibnamefont{Carbonara}},
  \bibnamefont{et~al.}, \bibinfo{journal}{Astropart. Phys.}
  \textbf{\bibinfo{volume}{28}}, \bibinfo{pages}{495} (\bibinfo{year}{2008}).

\bibitem[{\citenamefont{Dawson et~al.}(2005)\citenamefont{Dawson, Howard,
  Akimov, Araujo, Bewick, Davidge, Jones, Joshi, Lebedenko, Liubarsky
  et~al.}}]{Dawson:2005}
\bibinfo{author}{\bibfnamefont{J.~V.} \bibnamefont{Dawson}},
  \bibinfo{author}{\bibfnamefont{A.~S.} \bibnamefont{Howard}},
  \bibinfo{author}{\bibfnamefont{D.}~\bibnamefont{Akimov}},
  \bibinfo{author}{\bibfnamefont{H.}~\bibnamefont{Araujo}},
  \bibinfo{author}{\bibfnamefont{A.}~\bibnamefont{Bewick}},
  \bibinfo{author}{\bibfnamefont{D.~C.~R.} \bibnamefont{Davidge}},
  \bibinfo{author}{\bibfnamefont{W.~G.} \bibnamefont{Jones}},
  \bibinfo{author}{\bibfnamefont{M.}~\bibnamefont{Joshi}},
  \bibinfo{author}{\bibfnamefont{V.~N.} \bibnamefont{Lebedenko}},
  \bibinfo{author}{\bibfnamefont{I.}~\bibnamefont{Liubarsky}},
  \bibnamefont{et~al.}, \bibinfo{journal}{Nucl. Instrum. Meth. A}
  \textbf{\bibinfo{volume}{545}}, \bibinfo{pages}{690 } (\bibinfo{year}{2005}).

\bibitem[{\citenamefont{Kwong et~al.}(2010)\citenamefont{Kwong, Brusov, Shutt,
  Dahl, Bolozdynya, and Bradley}}]{Kwong:2010}
\bibinfo{author}{\bibfnamefont{J.}~\bibnamefont{Kwong}},
  \bibinfo{author}{\bibfnamefont{P.}~\bibnamefont{Brusov}},
  \bibinfo{author}{\bibfnamefont{T.}~\bibnamefont{Shutt}},
  \bibinfo{author}{\bibfnamefont{C.~E.} \bibnamefont{Dahl}},
  \bibinfo{author}{\bibfnamefont{A.~I.} \bibnamefont{Bolozdynya}},
  \bibnamefont{and} \bibinfo{author}{\bibfnamefont{A.}~\bibnamefont{Bradley}},
  \bibinfo{journal}{Nucl. Instrum. Meth. A} \textbf{\bibinfo{volume}{612}},
  \bibinfo{pages}{328 } (\bibinfo{year}{2010}).

\bibitem[{\citenamefont{Ueshima et~al.}(2011)\citenamefont{Ueshima, Abe,
  Hiraide, Hirano, Kishimoto, Kobayashi, Koshio, Liu, Martens, Moriyama
  et~al.}}]{Ueshima:2011}
\bibinfo{author}{\bibfnamefont{K.}~\bibnamefont{Ueshima}},
  \bibinfo{author}{\bibfnamefont{K.}~\bibnamefont{Abe}},
  \bibinfo{author}{\bibfnamefont{K.}~\bibnamefont{Hiraide}},
  \bibinfo{author}{\bibfnamefont{S.}~\bibnamefont{Hirano}},
  \bibinfo{author}{\bibfnamefont{Y.}~\bibnamefont{Kishimoto}},
  \bibinfo{author}{\bibfnamefont{K.}~\bibnamefont{Kobayashi}},
  \bibinfo{author}{\bibfnamefont{Y.}~\bibnamefont{Koshio}},
  \bibinfo{author}{\bibfnamefont{J.}~\bibnamefont{Liu}},
  \bibinfo{author}{\bibfnamefont{K.}~\bibnamefont{Martens}},
  \bibinfo{author}{\bibfnamefont{S.}~\bibnamefont{Moriyama}},
  \bibnamefont{et~al.}, \bibinfo{journal}{arXiv}  (\bibinfo{year}{2011}),
  \eprint{1106.2209v1}.

\bibitem[{\citenamefont{Gastler et~al.}(2011)\citenamefont{Gastler, Kearns,
  Klein, Hime, Lippincott, McKinsey, Nikkel, and Seibert}}]{Gastler:2009}
\bibinfo{author}{\bibfnamefont{D.}~\bibnamefont{Gastler}},
  \bibinfo{author}{\bibfnamefont{E.}~\bibnamefont{Kearns}},
  \bibinfo{author}{\bibfnamefont{J.}~\bibnamefont{Klein}},
  \bibinfo{author}{\bibfnamefont{A.}~\bibnamefont{Hime}},
  \bibinfo{author}{\bibfnamefont{W.~H.} \bibnamefont{Lippincott}},
  \bibinfo{author}{\bibfnamefont{D.~N.} \bibnamefont{McKinsey}},
  \bibinfo{author}{\bibfnamefont{J.~A.} \bibnamefont{Nikkel}},
  \bibnamefont{and} \bibinfo{author}{\bibfnamefont{S.}~\bibnamefont{Seibert}},
  \bibinfo{journal}{arXiv:1004.0373}  (\bibinfo{year}{2011}),
  \bibinfo{note}{submitted to Phys. Rev. C}.

\bibitem[{\citenamefont{Chepel et~al.}(2006)\citenamefont{Chepel, Solovov,
  Neves, Pereira, Mendes, Silva, Lindote, da~Cunha, Lopes, and
  Kossionides}}]{Chepel:2006}
\bibinfo{author}{\bibfnamefont{V.}~\bibnamefont{Chepel}},
  \bibinfo{author}{\bibfnamefont{V.}~\bibnamefont{Solovov}},
  \bibinfo{author}{\bibfnamefont{F.}~\bibnamefont{Neves}},
  \bibinfo{author}{\bibfnamefont{A.}~\bibnamefont{Pereira}},
  \bibinfo{author}{\bibfnamefont{P.~J.} \bibnamefont{Mendes}},
  \bibinfo{author}{\bibfnamefont{C.~P.} \bibnamefont{Silva}},
  \bibinfo{author}{\bibfnamefont{A.}~\bibnamefont{Lindote}},
  \bibinfo{author}{\bibfnamefont{J.~P.} \bibnamefont{da~Cunha}},
  \bibinfo{author}{\bibfnamefont{M.~I.} \bibnamefont{Lopes}}, \bibnamefont{and}
  \bibinfo{author}{\bibfnamefont{S.}~\bibnamefont{Kossionides}},
  \bibinfo{journal}{Astropart. Phys.} \textbf{\bibinfo{volume}{26}},
  \bibinfo{pages}{58} (\bibinfo{year}{2006}).

\bibitem[{\citenamefont{Aprile et~al.}(2005)\citenamefont{Aprile, Giboni,
  Majewski, Ni, Yamashita, Hasty, Manzur, and McKinsey}}]{Aprile:2005}
\bibinfo{author}{\bibfnamefont{E.}~\bibnamefont{Aprile}},
  \bibinfo{author}{\bibfnamefont{K.~L.} \bibnamefont{Giboni}},
  \bibinfo{author}{\bibfnamefont{P.}~\bibnamefont{Majewski}},
  \bibinfo{author}{\bibfnamefont{K.}~\bibnamefont{Ni}},
  \bibinfo{author}{\bibfnamefont{M.}~\bibnamefont{Yamashita}},
  \bibinfo{author}{\bibfnamefont{R.}~\bibnamefont{Hasty}},
  \bibinfo{author}{\bibfnamefont{A.}~\bibnamefont{Manzur}}, \bibnamefont{and}
  \bibinfo{author}{\bibfnamefont{D.~N.} \bibnamefont{McKinsey}},
  \bibinfo{journal}{Phys. Rev. D} \textbf{\bibinfo{volume}{72}},
  \bibinfo{pages}{72006} (\bibinfo{year}{2005}).

\bibitem[{\citenamefont{Aprile et~al.}(2009)\citenamefont{Aprile, Baudis, Choi,
  Giboni, Lim, Manalaysay, Monzani, Plante, Santorelli, and
  Yamashita}}]{Aprile:2009}
\bibinfo{author}{\bibfnamefont{E.}~\bibnamefont{Aprile}},
  \bibinfo{author}{\bibfnamefont{L.}~\bibnamefont{Baudis}},
  \bibinfo{author}{\bibfnamefont{B.}~\bibnamefont{Choi}},
  \bibinfo{author}{\bibfnamefont{K.~L.} \bibnamefont{Giboni}},
  \bibinfo{author}{\bibfnamefont{K.}~\bibnamefont{Lim}},
  \bibinfo{author}{\bibfnamefont{A.}~\bibnamefont{Manalaysay}},
  \bibinfo{author}{\bibfnamefont{M.~E.} \bibnamefont{Monzani}},
  \bibinfo{author}{\bibfnamefont{G.}~\bibnamefont{Plante}},
  \bibinfo{author}{\bibfnamefont{R.}~\bibnamefont{Santorelli}},
  \bibnamefont{and}
  \bibinfo{author}{\bibfnamefont{M.}~\bibnamefont{Yamashita}},
  \bibinfo{journal}{Phys. Rev. C} \textbf{\bibinfo{volume}{79}},
  \bibinfo{pages}{045807} (\bibinfo{year}{2009}).

\bibitem[{\citenamefont{Manzur et~al.}(2010)\citenamefont{Manzur, Curioni,
  Kastens, McKinsey, Ni, and Wongjirad}}]{Manzur:2010}
\bibinfo{author}{\bibfnamefont{A.}~\bibnamefont{Manzur}},
  \bibinfo{author}{\bibfnamefont{A.}~\bibnamefont{Curioni}},
  \bibinfo{author}{\bibfnamefont{L.}~\bibnamefont{Kastens}},
  \bibinfo{author}{\bibfnamefont{D.~N.} \bibnamefont{McKinsey}},
  \bibinfo{author}{\bibfnamefont{K.}~\bibnamefont{Ni}}, \bibnamefont{and}
  \bibinfo{author}{\bibfnamefont{T.}~\bibnamefont{Wongjirad}},
  \bibinfo{journal}{Phys. Rev. C} \textbf{\bibinfo{volume}{81}},
  \bibinfo{pages}{025808} (\bibinfo{year}{2010}).

\bibitem[{\citenamefont{Plante et~al.}(2011)\citenamefont{Plante, Aprile,
  Budnik, Choi, Giboni, Goetzke, Lang, Lim, and Fernandez}}]{Plante:2011}
\bibinfo{author}{\bibfnamefont{G.}~\bibnamefont{Plante}},
  \bibinfo{author}{\bibfnamefont{E.}~\bibnamefont{Aprile}},
  \bibinfo{author}{\bibfnamefont{R.}~\bibnamefont{Budnik}},
  \bibinfo{author}{\bibfnamefont{B.}~\bibnamefont{Choi}},
  \bibinfo{author}{\bibfnamefont{K.~L.} \bibnamefont{Giboni}},
  \bibinfo{author}{\bibfnamefont{L.~W.} \bibnamefont{Goetzke}},
  \bibinfo{author}{\bibfnamefont{R.~F.} \bibnamefont{Lang}},
  \bibinfo{author}{\bibfnamefont{K.~E.} \bibnamefont{Lim}}, \bibnamefont{and}
  \bibinfo{author}{\bibfnamefont{A.~J.~M.} \bibnamefont{Fernandez}},
  \bibinfo{journal}{arXiv}  (\bibinfo{year}{2011}), \eprint{1104.2587v1}.

\bibitem[{\citenamefont{McKinsey et~al.}(1997)\citenamefont{McKinsey, Brome,
  Butterworth, Golub, Habicht, Huffman, Lamoreaux, Mattoni, and
  Doyle}}]{McKinsey:1997}
\bibinfo{author}{\bibfnamefont{D.~N.} \bibnamefont{McKinsey}},
  \bibinfo{author}{\bibfnamefont{C.~R.} \bibnamefont{Brome}},
  \bibinfo{author}{\bibfnamefont{J.~S.} \bibnamefont{Butterworth}},
  \bibinfo{author}{\bibfnamefont{R.}~\bibnamefont{Golub}},
  \bibinfo{author}{\bibfnamefont{K.}~\bibnamefont{Habicht}},
  \bibinfo{author}{\bibfnamefont{P.~R.} \bibnamefont{Huffman}},
  \bibinfo{author}{\bibfnamefont{S.~K.} \bibnamefont{Lamoreaux}},
  \bibinfo{author}{\bibfnamefont{C.~E.~H.} \bibnamefont{Mattoni}},
  \bibnamefont{and} \bibinfo{author}{\bibfnamefont{J.~M.} \bibnamefont{Doyle}},
  \bibinfo{journal}{Nucl. Instrum. Meth. B} \textbf{\bibinfo{volume}{132}},
  \bibinfo{pages}{351 } (\bibinfo{year}{1997}).

\bibitem[{Nup()}]{Nupure}
{\bibinfo{title}{Nupure corporation}}, \bibinfo{note}{Omni Nupure III,
  www.nupure.com}.

\bibitem[{\citenamefont{Lippincott et~al.}(2010)\citenamefont{Lippincott, Cahn,
  Gastler, Kastens, Kearns, McKinsey, and Nikkel}}]{Lippincott:2010}
\bibinfo{author}{\bibfnamefont{W.~H.} \bibnamefont{Lippincott}},
  \bibinfo{author}{\bibfnamefont{S.~B.}~\bibnamefont{Cahn}},
  \bibinfo{author}{\bibfnamefont{D.}~\bibnamefont{Gastler}},
  \bibinfo{author}{\bibfnamefont{L.~W.}~\bibnamefont{Kastens}},
  \bibinfo{author}{\bibfnamefont{E.}~\bibnamefont{Kearns}},
  \bibinfo{author}{\bibfnamefont{D.~N.} \bibnamefont{McKinsey}},
  \bibnamefont{and} \bibinfo{author}{\bibfnamefont{J.~A.}
  \bibnamefont{Nikkel}}, \bibinfo{journal}{Phys. Rev. C}\textbf{\bibinfo{volume}{81}},
  \bibinfo{pages}{045803} (\bibinfo{year}{2010}).


\bibitem[{\citenamefont{Brun and Rademakers}(1997)}]{Brun:1997}
\bibinfo{author}{\bibfnamefont{R.}~\bibnamefont{Brun}} \bibnamefont{and}
  \bibinfo{author}{\bibfnamefont{F.}~\bibnamefont{Rademakers}},
  \bibinfo{journal}{Nucl. Instrum. Meth. A} \textbf{\bibinfo{volume}{389}},
  \bibinfo{pages}{81} (\bibinfo{year}{1997}).

\bibitem[{\citenamefont{Nikkel et~al.}(2007)\citenamefont{Nikkel, Lippincott,
  and McKinsey}}]{Nikkel:2007a}
\bibinfo{author}{\bibfnamefont{J.~A.} \bibnamefont{Nikkel}},
  \bibinfo{author}{\bibfnamefont{W.~H.} \bibnamefont{Lippincott}},
  \bibnamefont{and} \bibinfo{author}{\bibfnamefont{D.~N.}
  \bibnamefont{McKinsey}}, \bibinfo{journal}{J. Instrum.}
  \textbf{\bibinfo{volume}{2}}, \bibinfo{pages}{P11004} (\bibinfo{year}{2007}).

\bibitem[{The()}]{Thermo}
{\bibinfo{title}{Thermo electron}}, \bibinfo{note}{Model MP320,
  www.thermo.com}.

\bibitem[{RAT()}]{RAT}
\emph{\bibinfo{title}{Reactor analysis tool}},
  \bibinfo{note}{https://deapclean.org/rat/trac/}.

\bibitem[{\citenamefont{Agostinelli et~al.}(2003)\citenamefont{Agostinelli,
  Allison, Amako, Apostolakis, Araujo, Arce, Asai, Axen, Banerjee, Barrand
  et~al.}}]{Agostinelli:2003}
\bibinfo{author}{\bibfnamefont{S.}~\bibnamefont{Agostinelli}},
  \bibinfo{author}{\bibfnamefont{J.}~\bibnamefont{Allison}},
  \bibinfo{author}{\bibfnamefont{K.}~\bibnamefont{Amako}},
  \bibinfo{author}{\bibfnamefont{J.}~\bibnamefont{Apostolakis}},
  \bibinfo{author}{\bibfnamefont{H.}~\bibnamefont{Araujo}},
  \bibinfo{author}{\bibfnamefont{P.}~\bibnamefont{Arce}},
  \bibinfo{author}{\bibfnamefont{M.}~\bibnamefont{Asai}},
  \bibinfo{author}{\bibfnamefont{D.}~\bibnamefont{Axen}},
  \bibinfo{author}{\bibfnamefont{S.}~\bibnamefont{Banerjee}},
  \bibinfo{author}{\bibfnamefont{G.}~\bibnamefont{Barrand}},
  \bibnamefont{et~al.}, \bibinfo{journal}{Nucl. Instrum. Meth. A}
  \textbf{\bibinfo{volume}{506}}, \bibinfo{pages}{250} (\bibinfo{year}{2003}).

\bibitem[{\citenamefont{Acciarri
  et~al.}(2008{\natexlab{a}})\citenamefont{Acciarri, Antonello, Baibussinov,
  Baldo-Ceolin, Benetti, Calaprice, Calligarich, Cambiaghi, Canci, Carbonara
  et~al.}}]{Acciarri:2008}
\bibinfo{author}{\bibfnamefont{R.}~\bibnamefont{Acciarri}},
  \bibinfo{author}{\bibfnamefont{M.}~\bibnamefont{Antonello}},
  \bibinfo{author}{\bibfnamefont{B.}~\bibnamefont{Baibussinov}},
  \bibinfo{author}{\bibfnamefont{M.}~\bibnamefont{Baldo-Ceolin}},
  \bibinfo{author}{\bibfnamefont{P.}~\bibnamefont{Benetti}},
  \bibinfo{author}{\bibfnamefont{F.}~\bibnamefont{Calaprice}},
  \bibinfo{author}{\bibfnamefont{E.}~\bibnamefont{Calligarich}},
  \bibinfo{author}{\bibfnamefont{M.}~\bibnamefont{Cambiaghi}},
  \bibinfo{author}{\bibfnamefont{N.}~\bibnamefont{Canci}},
  \bibinfo{author}{\bibfnamefont{F.}~\bibnamefont{Carbonara}},
  \bibnamefont{et~al.}, \bibinfo{journal}{arXiv:0804.1217v1}
  (\bibinfo{year}{2008}{\natexlab{a}}).

\bibitem[{\citenamefont{Acciarri
  et~al.}(2008{\natexlab{b}})\citenamefont{Acciarri, Antonello, Baibussinov,
  Baldo-Ceolin, Benetti, Calaprice, Calligarich, Cambiaghi, Canci, Carbonara
  et~al.}}]{Acciarri:2008a}
\bibinfo{author}{\bibfnamefont{R.}~\bibnamefont{Acciarri}},
  \bibinfo{author}{\bibfnamefont{M.}~\bibnamefont{Antonello}},
  \bibinfo{author}{\bibfnamefont{B.}~\bibnamefont{Baibussinov}},
  \bibinfo{author}{\bibfnamefont{M.}~\bibnamefont{Baldo-Ceolin}},
  \bibinfo{author}{\bibfnamefont{P.}~\bibnamefont{Benetti}},
  \bibinfo{author}{\bibfnamefont{F.}~\bibnamefont{Calaprice}},
  \bibinfo{author}{\bibfnamefont{E.}~\bibnamefont{Calligarich}},
  \bibinfo{author}{\bibfnamefont{M.}~\bibnamefont{Cambiaghi}},
  \bibinfo{author}{\bibfnamefont{N.}~\bibnamefont{Canci}},
  \bibinfo{author}{\bibfnamefont{F.}~\bibnamefont{Carbonara}},
  \bibnamefont{et~al.}, \bibinfo{journal}{arXiv:0804.1222v1}
  (\bibinfo{year}{2008}{\natexlab{b}}).

\bibitem[{\citenamefont{Himi et~al.}(1982)\citenamefont{Himi, Takahashi, Ruan,
  and Kubota}}]{Himi:1982}
\bibinfo{author}{\bibfnamefont{S.}~\bibnamefont{Himi}},
  \bibinfo{author}{\bibfnamefont{T.}~\bibnamefont{Takahashi}},
  \bibinfo{author}{\bibfnamefont{J.}~\bibnamefont{Ruan}}, \bibnamefont{and}
  \bibinfo{author}{\bibfnamefont{S.}~\bibnamefont{Kubota}},
  \bibinfo{journal}{Nucl. Instrum. Methods} \textbf{\bibinfo{volume}{203}},
  \bibinfo{pages}{153} (\bibinfo{year}{1982}).

\bibitem[{\citenamefont{Lindhard et~al.}(1963)\citenamefont{Lindhard, Scharff,
  and Schiott}}]{Lindhard:1963}
\bibinfo{author}{\bibfnamefont{J.}~\bibnamefont{Lindhard}},
  \bibinfo{author}{\bibfnamefont{M.}~\bibnamefont{Scharff}}, \bibnamefont{and}
  \bibinfo{author}{\bibfnamefont{H.}~\bibnamefont{Schiott}},
  \bibinfo{journal}{Mat. Fys. Medd. Dan. Vid. Selsk.}
  \textbf{\bibinfo{volume}{33}}, \bibinfo{pages}{1} (\bibinfo{year}{1963}).

\bibitem[{\citenamefont{Benoit et~al.}(2007)\citenamefont{Benoit, Berg{\'e},
  Bl{\"u}mer, Broniatowski, Censier, Chantelauze, Chapellier, Chardin, Collin,
  Defay et~al.}}]{Benoit:2007}
\bibinfo{author}{\bibfnamefont{A.}~\bibnamefont{Benoit}},
  \bibinfo{author}{\bibfnamefont{L.}~\bibnamefont{Berg{\'e}}},
  \bibinfo{author}{\bibfnamefont{J.}~\bibnamefont{Bl{\"u}mer}},
  \bibinfo{author}{\bibfnamefont{A.}~\bibnamefont{Broniatowski}},
  \bibinfo{author}{\bibfnamefont{B.}~\bibnamefont{Censier}},
  \bibinfo{author}{\bibfnamefont{A.}~\bibnamefont{Chantelauze}},
  \bibinfo{author}{\bibfnamefont{M.}~\bibnamefont{Chapellier}},
  \bibinfo{author}{\bibfnamefont{G.}~\bibnamefont{Chardin}},
  \bibinfo{author}{\bibfnamefont{S.}~\bibnamefont{Collin}},
  \bibinfo{author}{\bibfnamefont{X.}~\bibnamefont{Defay}},
  \bibnamefont{et~al.}, \bibinfo{journal}{Nucl. Instrum. Meth. A}
  \textbf{\bibinfo{volume}{577}}, \bibinfo{pages}{558 } (\bibinfo{year}{2007}).

\bibitem[{\citenamefont{Hitachi}(2005)}]{Hitachi:2005}
\bibinfo{author}{\bibfnamefont{A.}~\bibnamefont{Hitachi}},
  \bibinfo{journal}{Astropart. Phys.} \textbf{\bibinfo{volume}{24}},
  \bibinfo{pages}{247 } (\bibinfo{year}{2005}).

\bibitem[{\citenamefont{Mei et~al.}(2008)\citenamefont{Mei, Yin, Stonehill, and
  Hime}}]{Mei:2008}
\bibinfo{author}{\bibfnamefont{D.-M.} \bibnamefont{Mei}},
  \bibinfo{author}{\bibfnamefont{Z.-B.} \bibnamefont{Yin}},
  \bibinfo{author}{\bibfnamefont{L.}~\bibnamefont{Stonehill}},
  \bibnamefont{and} \bibinfo{author}{\bibfnamefont{A.}~\bibnamefont{Hime}},
  \bibinfo{journal}{Astropart. Phys.} \textbf{\bibinfo{volume}{30}},
  \bibinfo{pages}{12 } (\bibinfo{year}{2008}).

\bibitem[{\citenamefont{Doke et~al.}(2002)\citenamefont{Doke, Hitachi, Kikuchi,
  Masuda, Okada, and Shibamura}}]{Doke:2002}
\bibinfo{author}{\bibfnamefont{T.}~\bibnamefont{Doke}},
  \bibinfo{author}{\bibfnamefont{A.}~\bibnamefont{Hitachi}},
  \bibinfo{author}{\bibfnamefont{J.}~\bibnamefont{Kikuchi}},
  \bibinfo{author}{\bibfnamefont{K.}~\bibnamefont{Masuda}},
  \bibinfo{author}{\bibfnamefont{H.}~\bibnamefont{Okada}}, \bibnamefont{and}
  \bibinfo{author}{\bibfnamefont{E.}~\bibnamefont{Shibamura}},
  \bibinfo{journal}{Jpn. J. Appl. Phys. 1} \textbf{\bibinfo{volume}{41}},
  \bibinfo{pages}{1538} (\bibinfo{year}{2002}).

\bibitem[{\citenamefont{Sorensen and Dahl}(2011)}]{Sorensen:2011}
\bibinfo{author}{\bibfnamefont{P.}~\bibnamefont{Sorensen}} \bibnamefont{and}
  \bibinfo{author}{\bibfnamefont{C.~E.} \bibnamefont{Dahl}},
  \bibinfo{journal}{Phys. Rev. D} \textbf{\bibinfo{volume}{83}},
  \bibinfo{pages}{063501} (\bibinfo{year}{2011}).

\bibitem[{\citenamefont{MacMullin et~al.}(2010)\citenamefont{MacMullin,
  Henning, Kidd, Tornow, and Howell}}]{MacMullin:2010}
\bibinfo{author}{\bibfnamefont{S.}~\bibnamefont{MacMullin}},
  \bibinfo{author}{\bibfnamefont{R.}~\bibnamefont{Henning}},
  \bibinfo{author}{\bibfnamefont{M.}~\bibnamefont{Kidd}},
  \bibinfo{author}{\bibfnamefont{W.}~\bibnamefont{Tornow}}, \bibnamefont{and}
  \bibinfo{author}{\bibfnamefont{C.}~\bibnamefont{Howell}},
  \bibinfo{journal}{Bulletin of the APS April Meeting}  (\bibinfo{year}{2010}).

\bibitem[{\citenamefont{Ziegler et~al.}(1985)\citenamefont{Ziegler, Biersack,
  and Littmark}}]{Ziegler:1985}
\bibinfo{author}{\bibfnamefont{J.~F.} \bibnamefont{Ziegler}},
  \bibinfo{author}{\bibfnamefont{J.~P.} \bibnamefont{Biersack}},
  \bibnamefont{and} \bibinfo{author}{\bibfnamefont{U.}~\bibnamefont{Littmark}},
  \emph{\bibinfo{title}{The Stopping and Range of Ions in Matter, Vol. 1}}
  (\bibinfo{year}{1985}).

\bibitem[{\citenamefont{McLachlan and Peel}(2000)}]{McLachlan:2000}
\bibinfo{author}{\bibfnamefont{G.~J.} \bibnamefont{McLachlan}}
  \bibnamefont{and} \bibinfo{author}{\bibfnamefont{D.}~\bibnamefont{Peel}},
  \emph{\bibinfo{title}{Finite mixture models}} (\bibinfo{year}{2000}).

\bibitem[{\citenamefont{Matusita}(1954)}]{Matusita:1954}
\bibinfo{author}{\bibfnamefont{K.}~\bibnamefont{Matusita}},
  \bibinfo{journal}{Ann. I. Stat. Math.} \textbf{\bibinfo{volume}{7}},
  \bibinfo{pages}{67 } (\bibinfo{year}{1954}).

\bibitem[{\citenamefont{Efron and Tibshirani}(1993)}]{Efron:1993}
\bibinfo{author}{\bibfnamefont{B.}~\bibnamefont{Efron}} \bibnamefont{and}
  \bibinfo{author}{\bibfnamefont{R.}~\bibnamefont{Tibshirani}},
  \emph{\bibinfo{title}{An introduction to the bootstrap}}
  (\bibinfo{year}{1993}).

\bibitem[{\citenamefont{Suemoto and Kanzaki}(1979)}]{Suemoto:1979}
\bibinfo{author}{\bibfnamefont{T.}~\bibnamefont{Suemoto}} \bibnamefont{and}
  \bibinfo{author}{\bibfnamefont{H.}~\bibnamefont{Kanzaki}},
  \bibinfo{journal}{J. Phys. Soc. Jpn.} \textbf{\bibinfo{volume}{46}},
  \bibinfo{pages}{1554} (\bibinfo{year}{1979}).

\bibitem[{\citenamefont{Michniak et~al.}(2002)\citenamefont{Michniak, Alleaume,
  McKinsey, and Doyle}}]{Michniak:2002}
\bibinfo{author}{\bibfnamefont{R.~A.} \bibnamefont{Michniak}},
  \bibinfo{author}{\bibfnamefont{R.}~\bibnamefont{Alleaume}},
  \bibinfo{author}{\bibfnamefont{D.~N.} \bibnamefont{McKinsey}},
  \bibnamefont{and} \bibinfo{author}{\bibfnamefont{J.~M.} \bibnamefont{Doyle}},
  \bibinfo{journal}{Nucl. Instrum. Meth. A} \textbf{\bibinfo{volume}{482}},
  \bibinfo{pages}{387 } (\bibinfo{year}{2002}).

\bibitem[{\citenamefont{Leichner}(1973)}]{Leichner:1973}
\bibinfo{author}{\bibfnamefont{P.~K.} \bibnamefont{Leichner}},
  \bibinfo{journal}{Phys. Rev. A} \textbf{\bibinfo{volume}{8}},
  \bibinfo{pages}{815} (\bibinfo{year}{1973}).

\bibitem[{\citenamefont{Oka et~al.}(1974)\citenamefont{Oka, Rao, Redpath, and
  Firestone}}]{Oka:1974}
\bibinfo{author}{\bibfnamefont{T.}~\bibnamefont{Oka}},
  \bibinfo{author}{\bibfnamefont{K.~V. S.~R.} \bibnamefont{Rao}},
  \bibinfo{author}{\bibfnamefont{J.~L.} \bibnamefont{Redpath}},
  \bibnamefont{and} \bibinfo{author}{\bibfnamefont{R.~F.}
  \bibnamefont{Firestone}}, \bibinfo{journal}{J. Chem. Phys.}
  \textbf{\bibinfo{volume}{61}}, \bibinfo{pages}{4740} (\bibinfo{year}{1974}).

\bibitem[{\citenamefont{McKinsey et~al.}(1999)\citenamefont{McKinsey, Brome,
  Butterworth, Dzhosyuk, Huffman, Mattoni, Doyle, Golub, and
  Habicht}}]{McKinsey:1999}
\bibinfo{author}{\bibfnamefont{D.~N.} \bibnamefont{McKinsey}},
  \bibinfo{author}{\bibfnamefont{C.~R.} \bibnamefont{Brome}},
  \bibinfo{author}{\bibfnamefont{J.~S.} \bibnamefont{Butterworth}},
  \bibinfo{author}{\bibfnamefont{S.~N.} \bibnamefont{Dzhosyuk}},
  \bibinfo{author}{\bibfnamefont{P.~R.} \bibnamefont{Huffman}},
  \bibinfo{author}{\bibfnamefont{C.~E.~H.} \bibnamefont{Mattoni}},
  \bibinfo{author}{\bibfnamefont{J.~M.} \bibnamefont{Doyle}},
  \bibinfo{author}{\bibfnamefont{R.}~\bibnamefont{Golub}}, \bibnamefont{and}
  \bibinfo{author}{\bibfnamefont{K.}~\bibnamefont{Habicht}},
  \bibinfo{journal}{Phys. Rev. A} \textbf{\bibinfo{volume}{59}},
  \bibinfo{pages}{200 } (\bibinfo{year}{1999}).

\bibitem[{\citenamefont{Keto et~al.}(1974)\citenamefont{Keto, Soley, Stockton,
  and Fitzsimmons}}]{Keto:1974a}
\bibinfo{author}{\bibfnamefont{J.~W.} \bibnamefont{Keto}},
  \bibinfo{author}{\bibfnamefont{F.~J.} \bibnamefont{Soley}},
  \bibinfo{author}{\bibfnamefont{M.}~\bibnamefont{Stockton}}, \bibnamefont{and}
  \bibinfo{author}{\bibfnamefont{W.~A.} \bibnamefont{Fitzsimmons}},
  \bibinfo{journal}{Phys. Rev. A} \textbf{\bibinfo{volume}{10}},
  \bibinfo{pages}{887} (\bibinfo{year}{1974}).

\bibitem[{\citenamefont{Woodworth and Moos}(1975)}]{Woodworth:1975}
\bibinfo{author}{\bibfnamefont{J.~R.} \bibnamefont{Woodworth}}
  \bibnamefont{and} \bibinfo{author}{\bibfnamefont{H.~W.} \bibnamefont{Moos}},
  \bibinfo{journal}{Phys. Rev. A} \textbf{\bibinfo{volume}{12}},
  \bibinfo{pages}{2455} (\bibinfo{year}{1975}).

\bibitem[{\citenamefont{K{\"o}ymen et~al.}(1990)\citenamefont{K{\"o}ymen, Tang,
  Zhao, Dunning, and Walters}}]{Koymen:1990}
\bibinfo{author}{\bibfnamefont{A.}~\bibnamefont{K{\"o}ymen}},
  \bibinfo{author}{\bibfnamefont{F.~C.} \bibnamefont{Tang}},
  \bibinfo{author}{\bibfnamefont{X.}~\bibnamefont{Zhao}},
  \bibinfo{author}{\bibfnamefont{F.~B.} \bibnamefont{Dunning}},
  \bibnamefont{and} \bibinfo{author}{\bibfnamefont{G.~K.}
  \bibnamefont{Walters}}, \bibinfo{journal}{Chem. Phys. Lett.}
  \textbf{\bibinfo{volume}{168}}, \bibinfo{pages}{405 } (\bibinfo{year}{1990}).

\bibitem[{\citenamefont{McKinsey et~al.}(2003)\citenamefont{McKinsey, Brome,
  Dzhosyuk, Golub, Habicht, Huffman, Korobkina, Lamoreaux, Mattoni, Thompson
  et~al.}}]{McKinsey:2003}
\bibinfo{author}{\bibfnamefont{D.~N.} \bibnamefont{McKinsey}},
  \bibinfo{author}{\bibfnamefont{C.~R.} \bibnamefont{Brome}},
  \bibinfo{author}{\bibfnamefont{S.~N.} \bibnamefont{Dzhosyuk}},
  \bibinfo{author}{\bibfnamefont{R.}~\bibnamefont{Golub}},
  \bibinfo{author}{\bibfnamefont{K.}~\bibnamefont{Habicht}},
  \bibinfo{author}{\bibfnamefont{P.~R.} \bibnamefont{Huffman}},
  \bibinfo{author}{\bibfnamefont{E.}~\bibnamefont{Korobkina}},
  \bibinfo{author}{\bibfnamefont{S.~K.} \bibnamefont{Lamoreaux}},
  \bibinfo{author}{\bibfnamefont{C.~E.~H.} \bibnamefont{Mattoni}},
  \bibinfo{author}{\bibfnamefont{A.~K.} \bibnamefont{Thompson}},
  \bibnamefont{et~al.}, \bibinfo{journal}{Phys. Rev. A}
  \textbf{\bibinfo{volume}{67}}, \bibinfo{pages}{062716}
  (\bibinfo{year}{2003}).

\bibitem[{\citenamefont{King and Voltz}(1966)}]{King:1966}
\bibinfo{author}{\bibfnamefont{T.~A.} \bibnamefont{King}} \bibnamefont{and}
  \bibinfo{author}{\bibfnamefont{R.}~\bibnamefont{Voltz}},
  \bibinfo{journal}{Proc. R. Soc. A} \textbf{\bibinfo{volume}{289}},
  \bibinfo{pages}{424} (\bibinfo{year}{1966}).

\bibitem[{\citenamefont{Bruschi et~al.}(1972)\citenamefont{Bruschi, Mazzi, and
  Santini}}]{Bruschi:1972}
\bibinfo{author}{\bibfnamefont{L.}~\bibnamefont{Bruschi}},
  \bibinfo{author}{\bibfnamefont{G.}~\bibnamefont{Mazzi}}, \bibnamefont{and}
  \bibinfo{author}{\bibfnamefont{M.}~\bibnamefont{Santini}},
  \bibinfo{journal}{Phys. Rev. Lett.} \textbf{\bibinfo{volume}{28}},
  \bibinfo{pages}{1504} (\bibinfo{year}{1972}).

\bibitem[{\citenamefont{Loveland et~al.}(1972)\citenamefont{Loveland, Comber,
  and Spear}}]{Loveland:1972}
\bibinfo{author}{\bibfnamefont{R.~J.} \bibnamefont{Loveland}},
  \bibinfo{author}{\bibfnamefont{P.~G.~L.} \bibnamefont{Comber}},
  \bibnamefont{and} \bibinfo{author}{\bibfnamefont{W.~E.} \bibnamefont{Spear}},
  \bibinfo{journal}{Phys. Lett. A} \textbf{\bibinfo{volume}{39}},
  \bibinfo{pages}{225 } (\bibinfo{year}{1972}).

\bibitem[{\citenamefont{Kuper}(1961)}]{Kuper:1961}
\bibinfo{author}{\bibfnamefont{C.~G.} \bibnamefont{Kuper}},
  \bibinfo{journal}{Phys. Rev.} \textbf{\bibinfo{volume}{122}},
  \bibinfo{pages}{1007} (\bibinfo{year}{1961}).

\bibitem[{\citenamefont{Jortner et~al.}(1965)\citenamefont{Jortner, Kestner,
  Rice, and Cohen}}]{Jortner:1965}
\bibinfo{author}{\bibfnamefont{J.}~\bibnamefont{Jortner}},
  \bibinfo{author}{\bibfnamefont{N.~R.} \bibnamefont{Kestner}},
  \bibinfo{author}{\bibfnamefont{S.~A.} \bibnamefont{Rice}}, \bibnamefont{and}
  \bibinfo{author}{\bibfnamefont{M.~H.} \bibnamefont{Cohen}},
  \bibinfo{journal}{J. Chem. Phys.} \textbf{\bibinfo{volume}{43}},
  \bibinfo{pages}{2614} (\bibinfo{year}{1965}).

\bibitem[{\citenamefont{Springett et~al.}(1967)\citenamefont{Springett, Cohen,
  and Jortner}}]{Springett:1967}
\bibinfo{author}{\bibfnamefont{B.~E.} \bibnamefont{Springett}},
  \bibinfo{author}{\bibfnamefont{M.~H.} \bibnamefont{Cohen}}, \bibnamefont{and}
  \bibinfo{author}{\bibfnamefont{J.}~\bibnamefont{Jortner}},
  \bibinfo{journal}{Phys. Rev.} \textbf{\bibinfo{volume}{159}},
  \bibinfo{pages}{183} (\bibinfo{year}{1967}).

\bibitem[{\citenamefont{O'Malley}(1963)}]{OMalley:1963}
\bibinfo{author}{\bibfnamefont{T.~F.} \bibnamefont{O'Malley}},
  \bibinfo{journal}{Phys. Rev.} \textbf{\bibinfo{volume}{130}},
  \bibinfo{pages}{1020} (\bibinfo{year}{1963}).

\bibitem[{\citenamefont{Springett et~al.}(1968)\citenamefont{Springett,
  Jortner, and Cohen}}]{Springett:1968}
\bibinfo{author}{\bibfnamefont{B.~E.} \bibnamefont{Springett}},
  \bibinfo{author}{\bibfnamefont{J.}~\bibnamefont{Jortner}}, \bibnamefont{and}
  \bibinfo{author}{\bibfnamefont{M.~H.} \bibnamefont{Cohen}},
  \bibinfo{journal}{J. Chem. Phys.} \textbf{\bibinfo{volume}{48}},
  \bibinfo{pages}{2720} (\bibinfo{year}{1968}).

\bibitem[{\citenamefont{Miyakawa and Dexter}(1969)}]{Miyakawa:1969}
\bibinfo{author}{\bibfnamefont{T.}~\bibnamefont{Miyakawa}} \bibnamefont{and}
  \bibinfo{author}{\bibfnamefont{D.~L.} \bibnamefont{Dexter}},
  \bibinfo{journal}{Phys. Rev.} \textbf{\bibinfo{volume}{184}},
  \bibinfo{pages}{166} (\bibinfo{year}{1969}).

\bibitem[{\citenamefont{Dennis et~al.}(1969)\citenamefont{Dennis, Durbin,
  Fitzsimmons, Heybey, and Walters}}]{Dennis:1969}
\bibinfo{author}{\bibfnamefont{W.~S.} \bibnamefont{Dennis}},
  \bibinfo{author}{\bibfnamefont{E.}~\bibnamefont{Durbin}},
  \bibinfo{author}{\bibfnamefont{W.~A.} \bibnamefont{Fitzsimmons}},
  \bibinfo{author}{\bibfnamefont{O.}~\bibnamefont{Heybey}}, \bibnamefont{and}
  \bibinfo{author}{\bibfnamefont{G.~K.} \bibnamefont{Walters}},
  \bibinfo{journal}{Phys. Rev. Lett.} \textbf{\bibinfo{volume}{23}},
  \bibinfo{pages}{1083} (\bibinfo{year}{1969}).

\bibitem[{\citenamefont{Hill et~al.}(1971)\citenamefont{Hill, Heybey, and
  Walters}}]{Hill:1971}
\bibinfo{author}{\bibfnamefont{J.~C.} \bibnamefont{Hill}},
  \bibinfo{author}{\bibfnamefont{O.}~\bibnamefont{Heybey}}, \bibnamefont{and}
  \bibinfo{author}{\bibfnamefont{G.~K.} \bibnamefont{Walters}},
  \bibinfo{journal}{Phys. Rev. Lett.} \textbf{\bibinfo{volume}{26}},
  \bibinfo{pages}{1213} (\bibinfo{year}{1971}).

\bibitem[{\citenamefont{Hickman and Lane}(1971)}]{Hickman:1971}
\bibinfo{author}{\bibfnamefont{A.~P.} \bibnamefont{Hickman}} \bibnamefont{and}
  \bibinfo{author}{\bibfnamefont{N.~F.} \bibnamefont{Lane}},
  \bibinfo{journal}{Phys. Rev. Lett.} \textbf{\bibinfo{volume}{26}},
  \bibinfo{pages}{1216} (\bibinfo{year}{1971}).

\bibitem[{\citenamefont{Coakley and McKinsey}(2004)}]{Coakley:2004}
\bibinfo{author}{\bibfnamefont{K.~J.} \bibnamefont{Coakley}} \bibnamefont{and}
  \bibinfo{author}{\bibfnamefont{D.~N.} \bibnamefont{McKinsey}},
  \bibinfo{journal}{Nucl. Instrum. Meth. A} \textbf{\bibinfo{volume}{522}},
  \bibinfo{pages}{504 } (\bibinfo{year}{2004}).


\end{thebibliography}
\end{document}